\documentclass[12pt]{article}
\usepackage{amsmath,amsthm}
\usepackage{graphicx,subcaption}
\usepackage{enumerate}
\usepackage{natbib}
\usepackage{animate}
\usepackage{booktabs}
\usepackage{url} 
\usepackage[]{changes}

\RequirePackage[OT1]{fontenc}
\RequirePackage{amsthm,amsmath,bm,hyperref}
\newcommand{\pkg}[1]{{\normalfont\fontseries{b}\selectfont #1}}

\usepackage{undertilde}
\usepackage{graphicx}
\usepackage{float}
\usepackage{amsthm}
\usepackage{amsfonts}
\usepackage{amsmath,mathtools}
\usepackage{comment}
\DeclareMathAlphabet\mathbfcal{OMS}{cmsy}{b}{n} 
\usepackage{multirow}
\newtheorem{theorem}{Theorem}[section]
\newtheorem{corollary} {Corollary}[section]
\newtheorem{lemma} {Lemma}[section]
\newtheorem{proposition} {Proposition}[section]
\newcommand{\innp}[2]{\left\langle #1, #2 \right \rangle} 
\newcommand{\abs}[1]{\lvert #1 \rvert} 
\newcommand{\norm}[1]{\| #1 \|} 

\usepackage{titlesec}

\titleformat{\paragraph}
{\normalfont\normalsize\bfseries}{\theparagraph}{1em}{}
\titlespacing*{\paragraph}
{0pt}{3.25ex plus 1ex minus .2ex}{1.5ex plus .2ex}

\begin{document}

\def\spacingset#1{\renewcommand{\baselinestretch}%
{#1}\small\normalsize} \spacingset{1}


\title{\bf Multivariate Functional Singular Spectrum Analysis Over Different Dimensional Domains}
\author{Jordan Trinka \\
Department of Mathematical and Statistical Sciences, \\Marquette University, USA\\
and \\
Hossein Haghbin \\
Department of Statistics, \\ Persian Gulf University, Iran\\
and\\
Mehdi Maadooliat \\
Department of Mathematical and Statistical Sciences, \\Marquette University, USA}
\date{}
\maketitle
\vspace{-.2in}

\begin{abstract}
In this work, we develop multivariate functional singular spectrum analysis (MFSSA) over different dimensional domains which is the functional extension of multivariate singular spectrum analysis (MSSA). In the following, we provide all of the necessary theoretical details supporting the work as well as the implementation strategy that contains the recipes needed for the algorithm. We provide a simulation study showcasing the better performance in reconstruction accuracy of a multivariate functional time series (MFTS) signal found using MFSSA as compared to other approaches and we give a real data study showing how MFSSA enriches analysis using intraday temperature curves and remote sensing images of vegetation. MFSSA is available for use through the \pkg{Rfssa} R package.

\end{abstract}
\noindent%
\vfill

\newpage
\spacingset{1.5} 
\section{Introduction} \label{sec:int}

A common problem in time series analysis is detection, extraction, and exploration of mean, seasonal, trend, and noise components in time series data. A technique known as singular spectrum analysis (SSA) has been developed as a nonparametric, exploratory method which can be used to identify such interesting components in ordinary time series where observations are scalars \citep{golyandina2001}. Often times, many variables are observed as a result of a single stochastic process and investigation of time series components can be made richer by performing a multivariate analysis of these vector observations. The MSSA algorithm is a technique that has seen success over its univariate SSA counterpart in decomposing a multidimensional time series into components if the covariates are moderately correlated \citep{golyandina2012}. MSSA also has been broken up into two approaches of vertical MSSA (VMSSA) and horizontal MSSA (HMSSA) where VMSSA involves the vertical stacking of univariate Hankel trajectory matrices while HMSSA works with the horizontal stacking of the same elements \citep{hassani2018}. Over the course of the last 15 years, MSSA has seen significant success in various areas of application see \cite{groth2011, golyandina2012, silva2018, hassani2019}. 

Functional data analysis embodies the evaluation and exploration of data that is comprised of functions such as curves or surfaces \citep{ramsay2005}. Functional PCA (FPCA) is a technique that is used to find the most informative directions in a time-independent collection of functional subjects \citep{ramsay2005}. Univariate Functional Singular Spectrum Analysis (FSSA) was developed by \citet{haghbin2019} as a novel technique that is used to decompose a time-dependent collection of functional subjects, known as a functional time series (FTS), into mean, seasonal, trend, and noise components. FSSA works to decompose a FTS in a similar fashion as SSA using a functional singular value decomposition (fSVD). This method was compared with other techniques of dimension reduction of a FTS including dynamic functional principal component analysis (DFPCA) \citep{hormann2012} and it was found that FSSA is the ideal approach in terms of reconstruction accuracy.

Multivariate functional data are observed when a stochastic process gives rise to multiple different functions over possibly different dimensional domains. Multivariate FPCA (MFPCA) was developed so that more than one variable of functional subjects could be included in the analysis. \citet{chiou2014} extended MFPCA to include a normalized approach which accounts for differences in degrees of variability in the covariates as well as differences in units. MFPCA was further extended by \citet{happ2018} to account for different dimensional domains so that one could perform dimension reduction on multivariate functional data that might be comprised of curves, surfaces, or any other finite dimensional domain altogether. A primary assumption of MFPCA is that the functional data are independent of time. With the goal of performing dimension reduction on a MFTS, one might conjecture to use FSSA on the covariates independently of one another but this fails to capture any cross-correlations between variables. MFSSA provides us a way to perform dimension reduction of a MFTS while capturing these cross-correlations to further enrich analysis and strengthen reconstruction accuracy of the true signal. In addition, MFSSA is developed, in the following, to handle functions taken over any finite dimensional domain. This can allow the user to explore relationships between time dependent curves, images, or any other hyperplane.

The rest of the paper is organized to first introduce the reader to MSSA, we then discuss the functional extension of MSSA known as MFSSA and how one can generalize MSSA into MFSSA by developing both horizontal MFSSA (HMFSSA) and vertical MFSSA (VMFSSA). We also show that VMFSSA solves the same problem as MFSSA using a unitary operator. We finish the paper by discussing a simulation study illustrating when MFSSA outperforms all other known methods in terms of reconstruction accuracy and a real data study where we use weather station intraday temperature curves and remote sensing images in a bivariate analysis to explore some of the more interesting qualities of MFTS data through the use of MFSSA. In supplementary material, we provide further interesting plots and animations for our real data study, we provide another real data study that uses surface reflectance density curves, we develop HMFSSA fully, and we provide proofs of all lemmas and propositions. In addition to all of this work, the MFSSA algorithm has been implemented in the \pkg{Rfssa} package and we also include a shiny app that can be launched from within the package allowing the user to explore the work with already loaded data or their own data.

\section{General Scheme of MSSA}\label{mssa}

MSSA is a type of SSA developed to analyze multivariate time series. The algorithm is broken up into two different approaches known as VMSSA and HMSSA. The MSSA algorithm consists of the following four steps:

\subsubsection*{MSSA I. Embedding}\label{emmssa}
Given $p$ univariate time series of length $N$, $\{y_i^{(j)}\}_{i=1,\dots,N}^{j=1,\dots,p}$, a multivariate time series can be considered as a series of length $N$ of $p$-tuples, $\vec{y}_{i}:=\left(y_i^{(1)},\dots,y_i^{(p)}\right) \in \mathbb{H}:=\mathbb{R}^{p}$, in the form of $\mathbf{y}_{N}:=(\vec{y}_{1},\dots,\vec{y}_{N})$. One may choose an integer $L$, where $L < \frac{N}{2}$, set $K=N-L+1$, and create the set of $L \times K$, univariate trajectory matrices, $\{\mathbf{X}^{\left(j\right)}\}_{j=1}^{p}$. These trajectory matrices have the form
\begin{equation}\label{unitrajmat}
\mathbf{X}^{\left(j\right)}:=\left[\mathbf{x}^{\left(j\right)}_{1},\dots,\mathbf{x}^{\left(j\right)}_{K}\right],
\end{equation}
\noindent where $\mathbf{x}_{k}^{\left(j\right)}:=\left[y_{k}^{\left(j\right)},\dots,y_{k+L-1}^{\left(j\right)}\right]^{\top}$ is referred as $k^{th}$ lagged vector associated with variable $j$. In the HMSSA, we concatenate the univariate trajectory matrices horizontally to obtain an $L \times pK$ multivariate trajectory matrix
\begin{equation}\label{hmssaem}
\mathbf{X}:= \left[\mathbf{X}^{\left(1\right)}, \dots, \mathbf{X}^{\left(p\right)}\right],
\end{equation}
\noindent where as in the VMSSA, we concatenate those univariate trajectory matrices vertically to obtain the associated $pL \times K$ multivariate trajectory matrix
\begin{equation}\label{vmssaem}
\mathbf{X}:= \begin{bmatrix}\mathbf{X}^{\left(1\right)}\\ \vdots\\ \mathbf{X}^{\left(p\right)}\end{bmatrix}.
\end{equation}
A Hankel matrix is defined as that whose antidiagonal elements are equivalent. One may note that since each univariate trajectory matrix, $\mathbf{X}^{(j)}$, is Hankel, therefore the multivariate trajectory matrix, $\mathbf{X}$, is block Hankel. 

As we shall see in Section \ref{vmfssa}, there would be an interchangeable relationship between the extension of VMSSA and MFSSA. Without loss of generality, in the remaining of this section we focus on the VMSSA. Therefore, we have that $\mathbf{X}:\mathbb{R}^{K} \rightarrow \mathbb{R}^{pL}$. Often times, this embedding step is viewed as applying an invertible transformation $\mathcal{T}:\mathbb{R}^{N} \rightarrow \mathbb{R}^{pL \times K}$ such that 
\begin{equation*}\label{Tmssa}
\mathbf{X}=\mathcal{T}(\mathbf{y}_{N}).
\end{equation*} 
\subsubsection*{MSSA II. Decomposition}\label{decmssa}
In the decomposition step we perform an SVD of the rank $r$ trajectory matrix, $\mathbf{X}$. The formulation for the SVD is given as
\begin{equation}
\mathbf{X}=\sum_{i=1}^{r}\sigma_{i}\mathbf{u}_{i}\mathbf{v}_{i}^{\top}=\sum_{i=1}^{r}\mathbf{X}_{i}, \nonumber
\end{equation}
\noindent where $\{\sigma_{i}\}_{i=1}^{r}$ are the singular values, $\{\mathbf{v}_{i}\}_{i=1}^{r}$ forms an orthonormal basis for the domain of $\mathbf{X}$, $\{\mathbf{u}_{i}\}_{i=1}^{r}$ forms an orthonormal basis for the range of $\mathbf{X}$, and the set of rank one matrices, $\{\mathbf{X}_{i}\}_{i=1}^{r}$, are known as elementary matrices.
\subsubsection*{MSSA III. Grouping}
For grouping, we partition the set of indices of $\{1,2,\dots,r\}$ into $m$ disjoint subsets $\{I_{1}, I_{2},\dots,I_{m} \}$ such that for any positive integer $q=1,\cdots,m$, the matrix $\mathbf{X}_{I_{q}}$ is defined as $\mathbf{X}_{I_{q}}:=\sum_{i \in I_{q}} \mathbf{X}_{i}$. This allows us to write the original trajectory matrix, $\mathbf{X}$, as
\begin{equation}\label{mssaexpansion}
\mathbf{X}=\mathbf{X}_{I_{1}}+\mathbf{X}_{I_{2}}+\cdots+\mathbf{X}_{I_{m}}.
\end{equation}
\noindent The grouping should be done so that each $\mathbf{X}_{I_{q}}$ describes a different feature of the original time series such as trend or seasonality which can be achieved by looking at exploratory plots like paired-plots or scree plots \citep{golyandina2001, hassani2018}.

\subsubsection*{MSSA IV. Reconstruction}
For any $pL\times K$ block Hankel matrix, one may use $\mathcal{T}^{-1}$ to obtain the associated multivariate time series. Note that the matrices $\mathbf{X}_{I_{q}}$'s ($q=1,\cdots,m$), given in \eqref{mssaexpansion}, are not necessary block Hankel, and therefore we can not use $\mathcal{T}^{-1}$ transformation. A popular remedy in the literature is to use orthogonal projection approach and approximate $\mathbf{X}_{I_{q}}$'s with appropriate block Hankel matrices. 

The matrix $\mathbf{X}_{I_{q}}$ can be written in the block form:
\begin{equation}\label{mssablockapprox}
\mathbf{X}_{I_{q}}= \begin{bmatrix}\mathbf{X}_{I_{q}}^{\left(1\right)}\\ \vdots\\ \mathbf{X}_{I_{q}}^{\left(p\right)}\end{bmatrix},\nonumber
\end{equation}
\noindent where $\mathbf{X}_{I_{q}}^{\left(j\right)}$ is an $L \times K$ matrix for $j=1,\dots,p$. The orthogonal projection of the $\mathbf{X}_{I_q}$ onto the space of the block Hankel matrices can be done by averaging the antidiagonal elements of each $\mathbf{X}_{I_{q}}^{\left(j\right)}$. We denote this approximated block Hankel matrix as $\widetilde{\mathbf{X}}_{I_{q}}$, and use the inverse transformation, $\mathcal{T}^{-1}$, to obtain
\begin{equation}
\tilde{\mathbf{y}}_{N}^{q}:=\mathcal{T}^{-1}(\widetilde{\mathbf{X}}_{I_{q}}), \nonumber
\end{equation}
\noindent and as such, we have $\mathbf{y}_{N}\approx\tilde{\mathbf{y}}_{N}^{1}+\dots+\tilde{\mathbf{y}}_{N}^{m}$.

\subsection{Separability}
Let $\mathbf{y}_{N}$ and $\mathbf{z}_{N}$ be two multivariate time series of length $N$. The weighted-correlation (w-correlation) between $y_{N}$ and $z_{N}$ is defined as
\begin{equation}\label{w-corr}
  \rho_{1,2}^{\left(w\right)}:= \frac{\innp{\mathbf{y}_{N}}{\mathbf{z}_{N}}_{w}}{\norm{\mathbf{y}_{N}}_{w}\norm{\mathbf{z}_{N}}_{w}}, \nonumber
\end{equation}
where $\innp{\mathbf{y}_{N}}{\mathbf{z}_{N}}_{w} := \sum_{j=1}^{p}\sum_{i=1}^{N}w_{i}y_{i}^{(j)}z_{i}^{(j)}$, $w_{i}:=\text{min}\{i,L,N-i+1\}$, and $\norm{\mathbf{y}}_{w}:=\sqrt{\innp{\mathbf{y}}{\mathbf{y}}_{w}}$. Like in all types of SSA, a correlation close to zero is desired for reconstructed time series.

\subsection{Parameter Selection}
The two parameters of SSA are the window length, $L$, and how one does the grouping. Since every type of SSA is a nonparametric, data-driven approach to analysis, differing choices of $L$ will give different results. A rule of thumb is that $L$ should be chosen to be a multiple of a periodicity that is present in the data but no greater than $N/2$ \citep{golyandina2001, golyandina2013}. As stated earlier, it is ideal to perform the grouping such that there is no correlation between reconstructions.

\section{Theoretical Foundations of MFSSA}\label{fssa-method}

The mathematical foundations in the following subsection are used throughout the paper and form the theoretical backbone of the MFSSA algorithm. 

\subsection{Preliminaries and Notations}
For each $j=1,\cdots,p$, consider an $m_j$-dimensional domain, ${{T}}_{j}$, to be a compact subset of $\mathbb{R}^{m_{j}}$, and let $\mathbb{F}_j:=L^{2}\left({{T}}_{j}\right)$ to be the Hilbert space of square integrable real functions defined on ${T}_{j}$. We define the Cartesian product space $\mathbb{H}:= \mathbb{F}_1 \times \cdots \times \mathbb{F}_{p}$, where each $\vec{x}\in \mathbb{H}$, can be denoted by the $p$-tuple $\left(x^{(1)},\dots,x^{(p)}\right)$. Note that $\mathbb{H}$ is a Hilbert space equipped with inner product
\begin{equation*}\label{observations inner product}
\innp{\vec{x}}{\vec{y}}_{\mathbb{H}}:=\sum_{j=1}^{p}\innp{x^{(j)}}{y^{(j)}}_{\mathbb{F}_j}=\sum_{j=1}^{p}\int_{{{T}}_{j}}x^{\left(j\right)}\left(s_{j}\right)y^{\left(j\right)}\left(s_{j}\right)ds_{j}, \quad s_{j} \in {{T}}_{j}, \nonumber
\end{equation*}
\noindent for some $\vec{x},\vec{y} \in \mathbb{H}$. We specify a MFTS of length $N$ as $\mathbf{y}_{N}:=(\vec{y}_{1},\dots,\vec{y}_{N})$, where $\vec{y}_{i} \in \mathbb{H}$.

Similarly, for a given $L\in\mathbb{N}$, $\mathbb{H}^L$ stands for the Cartesian product of $L$ copies of $\mathbb{H}$, and each $\mathbf{x}\in\mathbb{H}^L$ can be denoted by the $L$-tuple $\left(\vec{x}_{1},...,\vec{x}_{L}\right)$. Clearly $\mathbb{H}^L$ is a Hilbert space with respect to the inner product 
$$\innp{\mathbf{x}}{\mathbf{y}}_{\mathbb{H}^{L}}:=\sum_{i=1}^{L} \innp{\vec{x}_{i}}{\vec{y}_{i}}_{\mathbb{H}},\quad \mathrm{for}\ \mathbf{x},\mathbf{y} \in \mathbb{H}^{L}.$$

Next we define $\mathbb{H}^{L \times K}$ to be the space spanned by linear operators $\mathbfcal{V}:\mathbb{R}^{K} \rightarrow \mathbb{H}^{L}$, specified by $\left[\vec{v}_{i,k}\right]_{i=1,\dots,L}^{k=1,\dots,K}$, as
\begin{equation*}\label{rangeofy}
\mathbfcal{V}(\pmb{a}):=\left(\sum_{k=1}^{K}a_{k}\vec{v}_{1,k}, \dots,\sum_{k=1}^{K}a_{k}\vec{v}_{L,k}\right),\qquad \pmb{a}:=(a_1, a_2, \dots, a_K) \in \mathbb{R}^{K}, \nonumber
\end{equation*}
where $\vec{v}_{i,k}\in\mathbb{H}$. Now for two operators $\mathbfcal{V}$, $\mathbfcal{Z} \in \mathbb{H}^{L \times K}$, the Frobenius inner product can be defined as
\begin{equation*}\label{Frobenius Norm}
    \innp{\mathbfcal{V}}{\mathbfcal{Z}}_{F}:=\sum_{i=1}^{L}\sum_{k=1}^{K}\innp{\vec{v}_{i,k}}{\vec{z}_{i,k}}_{\mathbb{H}},\nonumber
\end{equation*}
\noindent which induces the Frobenius norm given by $\norm{\mathbfcal{V}}_{F}:=\sqrt{\innp{\mathbfcal{V}}{\mathbfcal{V}}_{F}}$. We denote by $\mathbb{H}_{H}^{L \times K}$ the Hankel subspace of $\mathbb{H}^{L \times K}$ such that for any $\widetilde{\mathbfcal{V}} = \left[\vec{\tilde{v}}_{i,k}\right] \in \mathbb{H}_{H}^{L \times K}$ there exists a $\vec{g}_{u} \in \mathbb{H}$ such that $\norm{\vec{\tilde{v}}_{i,k}-\vec{g}_{u}}_\mathbb{H}=0$ where $u=i+k$.

\subsection{MFSSA Algorithm}\label{mfssa.algorithm}
Similar to other SSA algorithms, MFSSA consists of four steps: Embedding, Decomposition, Grouping, and Reconstruction.

\subsubsection*{MFSSA I. Embedding}
As one may note the columns of a univariate trajectory matrix, as given in \eqref{unitrajmat}, are the corresponding lagged vectors. Therefore a trajectory matrix can be seen as a linear operator from $\mathbb{R}^K$ to the space of linear combinations of the lagged vectors. \citet{haghbin2019}  used this as a motivation to introduce the trajectory operator for FSSA. 

In a similar fashion, we define multivariate functional lagged vectors in $\mathbb{H}^{L}$ of the form
\begin{equation}\label{mlagvec}
\pmb{x}_{k}:=\left(\vec{y}_{k}, \vec{y}_{k+1},\dots,\vec{y}_{k+L-1}\right), \quad k=1,\dots,K.
\end{equation}
\noindent One may define a linear operator, specified with $\pmb{x}_k$'s, to obtain the trajectory operator, $\mathbfcal{X}:\mathbb{R}^{K} \rightarrow \mathbb{H}^{L}$. As such, for some $\pmb{a}=(a_1, a_2, \dots, a_K) \in \mathbb{R}^{K}$, we have
\begin{equation}\label{mfssa x op}
\mathbfcal{X}(\pmb{a}):=\sum_{k=1}^{K}a_{k}\pmb{x}_{k}.
\end{equation}
\noindent Notice that $\text{R}\left(\mathbfcal{X}\right)=\text{sp}\{\pmb{x}_{j}\}_{j=1}^{K}$ is the range of the operator $\mathbfcal{X}$ with rank $r$, where $1\leq r \leq \min(pL,K)$. This step of embedding can also be viewed as applying the invertible transformation, $\mathcal{T}:\mathbb{H}^{N} \rightarrow \mathbb{H}_{H}^{L \times K}$, such that 
\begin{equation}\label{B operator}
\mathbfcal{X}=\mathcal{T}(\mathbf{y}_{N}).
\end{equation}

\begin{proposition}\label{prop:traj}
 The operator given in \eqref{mfssa x op} is a bounded and linear operator with adjoint $\mathbfcal{X}^{*}:\mathbb{H}^{L} \rightarrow \mathbb{R}^{K}$
\begin{equation*}
\mathbfcal{X}^{*}\mathbf{z}:=\left(\innp{\pmb{x}_{1}}{\mathbf{z}}_{\mathbb{H}^{L}}, \innp{\pmb{x}_{2}}{\mathbf{z}}_{\mathbb{H}^{L}}, \dots, \innp{\pmb{x}_{K}}{\mathbf{z}}_{\mathbb{H}^{L}}\right)^{\top} \in \mathbb{R}^{K}.
\end{equation*}
\end{proposition}

\subsubsection*{MFSSA II. Decomposition}\label{mfssadecomp}
Notice that the compact operator $\mathbfcal{X}$, is of rank $r$. Therefore one may employ Theorem 7.6 of \cite{weidmann1980} and obtain the SVD for the operator $\mathbfcal{X}$ as
\begin{equation}\label{fsvd}
\mathbfcal{X}(\pmb{a})=\sum_{i=1}^{r}\sigma_{i}\langle \mathbf{v}_{i},\pmb{a}\rangle_{\mathbb{R}^K}\boldsymbol{\psi}_{i}=\sum_{i=1}^{r}\sigma_{i}\mathbf{v}_{i}\otimes \boldsymbol{\psi}_{i}\left(\pmb{a}\right)=\sum_{i=1}^{r} \mathbfcal{X}_{i}(\pmb{a}).
\end{equation}
\noindent Here, $\{\sigma_{i}\}_{i=1}^{r}$ are the singular values, $\{\mathbf{v}_{i}\}_{i=1}^{r}$ are the orthonormal right singular vectors spanning $\mathbb{R}^{r}$, $\{\boldsymbol{\psi}_{i}\}_{i=1}^{r}$ are the orthonormal left singular functions spanning an $r$-dimensional subspace of $\mathbb{H}^{L}$. Now we define the rank one elementary operators $\mathbfcal{X}_{i}:=\sigma_{i}\mathbf{v}_{i}\otimes \boldsymbol{\psi}_{i}$, where $\otimes$ stands for the tensor(outer) product. It is easy to see that $\mathbfcal{X}=\sum_i\mathbfcal{X}_i$. We call the result of \eqref{fsvd} the multivariate fSVD (mfSVD) of $\mathbfcal{X}$ and we call the set $(\sigma_{i},\boldsymbol{\psi}_{i},\mathbf{v}_{i})$ to be the $i^{th}$ eigentriple of $\mathbfcal{X}$.

\begin{proposition}\label{prop:mfssadecomp}
Let $(\sigma_{i},\boldsymbol{\psi}_{i},\mathbf{v}_{i})$ be the $i^{th}$ eigentriple of $\mathbfcal{X}$, $i=1,\dots,r$. The following hold:
\begin{equation}
\boldsymbol{\psi}_{i}=\sigma_{i}^{-1}\mathbfcal{X}\mathbf{v}_{i}, \quad \mathbf{v}_{i}=\sigma_{i}^{-1}\mathbfcal{X}^{*}\boldsymbol{\psi}_{i}. \nonumber
\end{equation}
\end{proposition}

\subsubsection*{MFSSA III. Grouping}
The grouping step of MFSSA follows the same flavor as the grouping step of MSSA. We partition the set of indices $\{1,2,\dots,r\}$ into $m$ disjoint subsets $\{I_{1},I_{2},\dots,I_{m}\}$ such that for any $q$, the operator $\mathbfcal{X}_{I_{q}}$ is defined as $\mathbfcal{X}_{I_{q}}:=\sum_{i \in I_{q}} \mathbfcal{X}_{i}$. As such, we write
\begin{equation}\label{mfssa traj op}
\mathbfcal{X} = \mathbfcal{X}_{I_{1}} + \mathbfcal{X}_{I_{2}} + \cdots + \mathbfcal{X}_{I_{m}}.\nonumber
\end{equation}
\noindent Similar to \cite{haghbin2019}, exploratory plots, such as scree plots, paired-plots, w-correlation plots, and others can be developed to determine how to obtain the $m$ disjoint groups.

\subsubsection*{MFSSA IV. Reconstruction}\label{mfssa:recon}
Let $\mathbfcal{Y} \in \mathbb{H}^{L \times K}$, then since $\mathbb{H}_{H}^{L \times K}$ is a closed subspace of $\mathbb{H}^{L \times K}$, we have by the Projection Theorem that there exists a unique $\widetilde{\mathbfcal{Y}} \in \mathbb{H}_{H}^{L \times K}$ such that
\begin{equation}\label{reconstruct XIq}
\norm{\mathbfcal{Y}-\widetilde{\mathbfcal{Y}}}_{F}^{2} \leq \norm{\mathbfcal{Y}-\widetilde{\mathbfcal{Z}}}_{F}^{2}, \nonumber
\end{equation}
\noindent for any $\widetilde{\mathbfcal{Z}} \in \mathbb{H}_{H}^{L \times K}$. Define the projector $\Pi:\mathbb{H}^{L \times K} \rightarrow \mathbb{H}_{H}^{L \times K}$ such that we have $\Pi \mathbfcal{Y}=\widetilde{\mathbfcal{Y}}$. We achieve this projection by using Lemma 3.1 of \cite{haghbin2019} and the resulting diagonal averaging technique that
\begin{equation}\label{fssadiagavg}
\vec{\tilde{y}}_{i,k}:=\frac{1}{n_{u}}\sum_{\left(n,m\right):n+m=u}\vec{y}_{n,m},
\end{equation}
\noindent where $n_{u}$ is the number of $\left(n,m\right)$ pairs such that $n+m=u$. With this projection, we have that $\Pi\mathbfcal{X}_{I_{q}}=\widetilde{\mathbfcal{X}}_{I_{q}}$ for $q=1,\dots,m$. We then employ the inverse of $\mathcal{T}$ from \eqref{B operator} to obtain the following formula for the reconstruction
\begin{equation}\label{mfssarecon}
\tilde{\mathbf{y}}^{q}_{N}:=\mathcal{T}^{-1}\widetilde{\mathbfcal{X}}_{I_{q}},\nonumber
\end{equation}
\noindent where $\mathbf{y}_{N} = \sum_{q=1}^{m}\tilde{\mathbf{y}}^{q}_{N}$.

\subsection{Separability}
Let $\mathbf{x}_{N} = \mathbf{y}_{N}+\mathbf{z}_{N}$ where each $\mathbf{y}_{N}$ and $\mathbf{z}_{N}$ are multivariate functional time series. We define the weighted-covariance between multivariate functional time series as
\begin{equation}\label{w-corr}
   \innp{\mathbf{y}_{N}}{\mathbf{z}_{N}}_{w} := \sum_{j=1}^{p}\sum_{i=1}^{N}w_{i}\innp{y_{i}^{(j)}}{z_{i}^{(j)}}_{\mathbb{F}_j},\nonumber
\end{equation}
\noindent where $w_{i} := \text{min}\{i, L, N-i+1\}$. We call $\mathbf{x}_{N}$ separable if $\innp{\mathbf{y}_{N}}{\mathbf{z}_{N}}_{w} = 0$. The weighted-covariance measure shown here can also be used to form a so-called $w$-correlation between MFTS.

\section{MFSSA Implementation}\label{mfssaimp}

Similar to the discussion of implementation in \cite{haghbin2019}, we observe discrete samples of functional data that are then converted into functional objects using smoothing methods. Techniques that are used to form the functional data observations can be found in \cite{ramsay2007}. Let $\{\nu_{i}^{(j)}\}_{i \in \mathbb{N}}$ be the collection of basis functions in $\mathbb{F}_j$ for $j = 1,...,p$. Each observation in $\mathbb{F}_j$ can be projected onto the subspace $F_{j}:=\text{sp}\{\nu_{i}^{(j)}\}_{i=1}^{d_{j}}$ where $d_{j}$ can be determined by a variety of techniques like cross-validation. To this end, each $y_{i}^{(j)} \in \mathbb{F}_{j}$ can be projected to $F_{j}$ as
\begin{equation}\label{basis expansion}
    \hat{y}_{i}^{(j)}:=\sum_{k=1}^{d_{j}}c_{i,k}^{(j)}\nu_{k}^{(j)}, \quad i=1,...,N, \quad c_{i,k}^{(j)} \in \mathbb{R}.\nonumber
\end{equation}
\noindent where $\hat{y}_{i}^{(j)} \in F_{j}$. Now we set $d_0:=0$, $d:=\sum_{j=0}^{p}d_{j}$, and $\mathbb{H}_d:=F_1\times \ldots \times F_p \subseteq \mathbb{H}.$ For the rest of this section we provide the implementation of the MFSSA on MFTS $\mathbf{y}_{N}:=(\vec{\hat{y}}_{1},\dots,\vec{\hat{y}}_{N})$, where $\vec{\hat{y}}_{i}:=(\hat{y}_{i}^{(1)},\ldots, \hat{y}_{i}^{(p)})\in \mathbb{H}_d$. 

For each $q\in\{1,\ldots, d\}$, there exist a unique $j_q\in\{1,\ldots,p\}$ such that $\sum_{i=0}^{j_q-1}d_i +1 \leq q \leq \sum_{i=0}^{j_q}d_i$. Now consider $\vec{\nu}_q \in \mathbb{H}_d$, as a multivariate functional object of length $p$ with all zero functions, except $j_q$-th element, which is $\nu_{\ell_q}^{(j_q)}$, where $\ell_q:={q-\sum_{i=0}^{j_q-1}d_i}$.

\begin{lemma}\label{lem:nu} The following holds:
\begin{itemize}
\item[i)] Each multivariate functional object $\vec{\hat{y}}_{i}$ can be uniquely represented as a linear combination of $\vec{\nu}_q$'s
\begin{equation}\label{vbasis expansion}
    \vec{\hat{y}}_{i}:=\sum_{q=1}^{d}c_{i,\ell_q}^{(j_q)}\vec{\nu}_q, \quad i=1,...,N.\nonumber
\end{equation}
\item[ii)] The set $\{\vec{\nu}_q\}_{q=1}^{d}$ is a basis system of $\mathbb{H}_{d}$.
\end{itemize}
\end{lemma}

Now for each $k\in\{1,\ldots, Ld\}$, one can see that there exist unique $q_k\in\{1,\ldots,d\}$ and $r_k\in\{1,\ldots,L\}$ such that $k=(q_k-1)L+r_k$. Consider ${\pmb \phi}_{k}$ as a functional vector of length $L$ with all zero functions, except $r_k$-th element, which is $\vec{\nu}_{q_k}$.

\begin{lemma}\label{lem:phi}
The sequence $\{{\pmb \phi}_{k}\}_{k=1}^{Ld}$ is a basis system for $\mathbb{H}_d^L$, where $\mathbb{H}_d^L$ is the Cartesian product of $L$ copies of $\mathbb{H}_d$.
\end{lemma}

Using the Lemma \ref{lem:phi}, one may define a linear operator $\mathbfcal{P}:\mathbb{R}^{Ld}\rightarrow \mathbb{H}_d^L$, specified with ${\pmb \phi}_{k}$'s, where each $\pmb{x}\in \mathbb{H}_d^L$ can be written as
\begin{equation*}\label{A}
	\pmb{x} = \sum_{i=1}^{Ld} b_{i}{\pmb \phi}_{i}=\mathbfcal{P}(\mathbf{b}).
\end{equation*}
We call ${\mathbf{b}}=(b_1, \ldots,b_{Ld})\in\mathbb{R}^{Ld},$ the corresponding coefficient vector of $\pmb{x}$ with respect to the operator $\mathbfcal{P}$. Similar to \eqref{mlagvec} one may define the functional lagged vectors for the MFTS $\mathbf{y}_{N}$ as $\pmb{x}_{k}=\left(\vec{\hat{y}}_{k},\vec{\hat{y}}_{k+1},\dots,\vec{\hat{y}}_{k+L-1}\right)\in \mathbb{H}_d^L$, where $k=1,\dots,K$. Therefore the associated trajectory operator, given in \eqref{mfssa x op}, would be $\mathbfcal{X}:\mathbb{R}^{K} \rightarrow \mathbb{H}_d^{L}$.

\begin{lemma}\label{lem:p} The following holds:
\begin{itemize}
\item[i)] The corresponding coefficient vector of the functional lagged vector $\pmb{x}_k$ with respect to the operator $\mathbfcal{P}$ is 
\begin{equation*}\label{lem:phi_b}
	{\bf b}_k:=\left[c_{k,\ell_1}^{(j_1)},\ldots, c_{k+L-1,\ell_1}^{(j_1)}, c_{k,\ell_2}^{(j_2)},\ldots, c_{k+L-1,\ell_2}^{(j_2)},\ldots, c_{k+L-1,\ell_d}^{(j_d)} \right]^\top\in \mathbb{R}^{Ld}.
\end{equation*}
\item[ii)] For any $\pmb{a} \in \mathbb{R}^{K}$, we have $\mathbfcal{X}(\pmb{a})=\mathbfcal{P}(\mathbf{B}\pmb{a}),$ where ${\mathbf B}:=\left[b_{k,i}\right]_{i=1,\dots,Ld}^{k=1,\dots,K}=\left[\mathbf{b}_{1}, \mathbf{b}_{2}, \dots, \mathbf{b}_{K}\right]_{Ld \times K}$, and $b_{k,i}$ is the $i^{th}$ element of ${\mathbf b}_{k}$.
\end{itemize}
\end{lemma}

The following theorem gives us the recipes necessary to obtain the eigentriples of $\mathbfcal{X}$.

\begin{theorem}\label{thm:recipe}
Suppose ${\mathbf X}:={\mathbf G}^{1/2}{\mathbf B}$ where ${\mathbf G}:=\left[\innp{\boldsymbol{\phi}_{i}}{\boldsymbol{\phi}_{j}}_{\mathbb{H}^{L}}\right]_{i,j=1}^{Ld}$ is the Gram matrix. Denote the collection $\left({\sigma}_{i},{\mathbf{v}}_{i},{\pmb{u}}_{i}\right)$ as the $i^{\text{th}}$ eigentriple of $\mathbf{X}$. Now define ${\boldsymbol{\psi}}_{i}:=\mathbfcal{P} ({\mathbf G}^{-1/2}{\pmb{u}}_{i})$. The following holds:
\begin{itemize}
\item[i)]$\mathbfcal{X}^{*}{\boldsymbol{\psi}}_{i}={\sigma}_{i}{\mathbf{v}}_{i}$
\item[ii)]$\mathbfcal{X}{\mathbf{v}}_{i}={\sigma}_{i}{\boldsymbol{\psi}}_{i}$
\item[iii)]The collection $\{{\boldsymbol{\psi}}_{i}\}_{i=1}^{r}$ form an orthonormal basis for $R(\mathbfcal{X})$.
\end{itemize}
\end{theorem}

\begin{corollary}
The collection of triples $({\sigma}_{i},{\mathbf{v}}_{i},{\boldsymbol{\psi}}_{i})_{i=1}^{r}$ defines the mfSVD of $\mathbfcal{X}$.
\end{corollary}

\section{Generalizing MSSA to MFSSA} 

One may note that a key step in extending different SSA approaches, is how to obtain the trajectory matrix (operator) in the embedding step (see e.g., Sections \ref{emmssa} and \ref{mfssa.algorithm}). Despite the fact in SSA, where the trajectory matrix is a linear combination of the associated lagged vectors, that is not the case for MSSA.

In Section 3, we obtain MFSSA by generalizing FSSA, where we introduce the trajectory operator as a linear combination of multivariate lagged vectors. Alternatively, one may mimic the approach of MSSA algorithms (HMSSA or VMSSA) and develop new trajectory operators that are not necessarily based on lagged vectors. The following subsections would extend HMSSA and VMSSA to obtain the functional versions respectively.

\subsection{From HMSSA to HMFSSA}\label{HMSSA2MMFSSA}
As one may see the columns of $\mathbf{X}^{(j)}$ in \eqref{unitrajmat}, $\mathbf{x}_k^{(j)}$'s, are the univariate lagged vectors for the $j^{th}$ variable. Therefore one can see the $\mathbf{X}^{(j)}$ as an operator from $\mathbb{R}^K \rightarrow \mathbb{R}^L$, which can be seen as a linear combination of these lagged vectors:
\begin{equation*}
\mathbf{X}^{(j)}\pmb{a}^{(j)}=\sum_{k=1}^{K}a_k^{(j)}\mathbf{x}_k^{(j)},\qquad \pmb{a}^{(j)}:=(a_1^{(j)},\dots,a_K^{(j)})\in \mathbb{R}^K.
\end{equation*}
In the embedding step of HMSSA, the trajectory matrix, given in \eqref{hmssaem}, can be seen as a linear operator, $\mathbf{X}:\mathbb{R}^{pK}\rightarrow \mathbb{R}^L$, where
\begin{equation}\label{hmssa:linear comb}
\mathbf{X}\pmb{a}=\sum_{j=1}^{p}\sum_{k=1}^{K}a_k^{(j)}\mathbf{x}_k^{(j)},\qquad \pmb{a}:=(\pmb{a}^{(1)},\dots,\pmb{a}^{(p)})\in \mathbb{R}^{pK}.
\end{equation}
In order to extend to the functional space, we need to assume that the lag vectors in HMFSSA, denoted with $\pmb{x}_k^{(j)}$, are in the space $\mathbb{F}_j^L$, for $j=1,\dots,p$. But the linear combination of $\pmb{x}_k^{(j)}$'s are well-defined if and only if $\mathbb{F}_1^L=\cdots=\mathbb{F}_p^L$, or equivalently $T_1=\dots=T_p$. We shall call the extension of this special case as HMFSSA and we present it in the supplementary material.

\subsection{From VMSSA to VMFSSA}\label{vmfssa}
In the embedding step of VMSSA, the trajectory matrix, given in \eqref{vmssaem}, can be seen as a linear operator, $\mathbf{X}:\mathbb{R}^{K}\rightarrow \mathbb{R}^{pL}$, with 
\begin{equation*}
\mathbf{X}\pmb{a}=\sum_{k=1}^{K}a_j\underline{\mathbf{x}}_k,\qquad \pmb{a}:=({a}_{1},\dots,{a}_{K})\in \mathbb{R}^{K}\quad\mathrm{and\ } \underline{\mathbf{x}}_k:=
\begin{bmatrix}\mathbf{x}_k^{(1)}\\ \vdots\\ \mathbf{x}_k^{(p)}\end{bmatrix}\in \mathbb{R}^{pL}.
\end{equation*}
\noindent To develop VMFSSA, we need to extend this operator to the functional space, i.e., $\underline{\mathbf{x}}_k$ should belong to a new unfolded Hilbert space,
$\mathbb{H}^{p,L}:=\underbrace{\mathbb{F}_{1}\times \cdots \times \mathbb{F}_{1}}_{L\text{ times}}\times \ldots \times \underbrace{\mathbb{F}_{p} \times \cdots \times \mathbb{F}_{p}}_{L\text{ times}}$. Here, each $\underline{\mathbf{x}}\in \mathbb{H}^{p,L}$ is denoted by \newline $\underline{\mathbf{x}}:=\left(x_{1}^{\left(1\right)}, \cdots, x_{L}^{\left(1\right)}, \ldots,  x_{1}^{\left(p\right)}, \cdots,  x_{L}^{\left(p\right)}\right)$. It is easy to see that $\mathbb{H}^{p,L}$ is a Hilbert space equipped with inner product
\begin{equation}\label{eqinnp}
\innp{\underline{\mathbf{x}}}{\underline{\mathbf{y}}}_{\mathbb{H}^{p,L}}:=\sum_{i=1}^{L}\sum_{j=1}^{p}\innp{x_{i}^{\left(j\right)}}{y_{i}^{\left(j\right)}}_{\mathbb{F}_{j}}=\sum_{i=1}^{L} \innp{\vec{x}_{i}}{\vec{y}_{i}}_{\mathbb{H}}=\innp{\mathbf{x}}{\mathbf{y}}_{\mathbb{H}^{L}}.\nonumber
\end{equation}
\noindent Therefore, there exists a unitary operator $\mathcal{U}:\mathbb{H}^{L} \rightarrow \mathbb{H}^{p,L}$ where $\mathcal{U}(\mathbf{x})=\underline{\mathbf{x}}$, and we have an isomorphism between $\mathbb{H}^{L}$ and $\mathbb{H}^{p,L}$. Now one may define the linear operator $\underline{\mathcal{X}}:\mathbb{R}^K\rightarrow\mathbb{H}^{p,L}$, specified with $\underline{\mathbf{x}}_k$'s, as 
\begin{equation*}
\underline{\mathcal{X}}\pmb{a}:=\sum_{k=1}^{K}a_j\underline{\mathbf{x}}_k,\qquad \pmb{a}\in \mathbb{R}^{K}\quad\mathrm{and}\quad \underline{\mathbf{x}}_k\in\mathbb{H}^{p,L}.
\end{equation*}

The following theorem illustrates the equivalency between the MFSSA and VMFSSA results.

\begin{theorem}\label{thm:vmfssa}
Let $({\sigma}_{i},{\mathbf{v}}_{i},\boldsymbol{\psi}_{i})_{i=1}^r$ to be the eigentriples of ${\mathbfcal{X}}$. The following holds:
\begin{itemize}
\item[i)] $\underline{\mathcal{X}}=\mathcal{U}\mathbfcal{X}$. 
\item[ii)]Furthermore, $\underline{\mathcal{X}}$ is a rank $r$ operator with the eigentriples $({\sigma}_{i},{\mathbf{v}}_{i},\underline{\boldsymbol{\psi}}_{i})_{i=1}^r$, where $\underline{\boldsymbol{\psi}}_{i}=\mathcal{U}\boldsymbol{\psi}_{i}$.
\end{itemize}
\end{theorem}

\noindent Therefore the decompositions obtained via MFSSA and VMFSSA are interchangeable and subsequently the respective groupings and reconstructions are equivalent. 

\section{Numerical Studies}

In order to explore the capabilities of MFSSA and HMFSSA we implement a simulation study where we compare our two novel algorithms to other approaches of MFTS reconstruction of the true signal. We also present an application to remote sensing data which is used to further illustrate the interesting qualities of MFTS data that are discovered by MFSSA.

\subsection{Simulation Study}\label{simstudy}
For the simulation, we generate a bivariate FTS of lengths $N=\{100,200\}$ by projecting the following discrete observations sampled in equidistance on the unit interval onto a B-spline basis with 15 degrees of freedom
\begin{align*}
Y_{t}^{\left(1\right)}\left(s_{i}\right)&:=y_{t}^{\left(1\right)}+X_{t}^{\left(1\right)}\\
Y_{t}^{\left(2\right)}\left(s_{i}\right)&:=y_{t}^{\left(2\right)}+X_{t}^{\left(2\right)},\quad s_{i} \in \left[0,1\right],\quad i=1,\dots,100,\quad t=1,\dots,N.
\end{align*}
\noindent where $y_{t}^{\left(1\right)}:=\mu_{t}+\delta_{t}^{\left(1\right)}$ and $y_{t}^{\left(2\right)}:=\delta_{t}^{\left(2\right)}$ are nonrandom, true signal terms. We take $\mu_{t}:=kt$ as an increasing trend component with $k=\{0.00,0.02\}$, $\delta_{t}^{\left(j\right)}$ are taken as seasonal components with expressions given as
\begin{eqnarray*}
\delta_{t}^{\left(1\right)}&:=&e^{s_{i}^{2}}\cos\left(2\pi\omega_{1}t\right)-e^{1-s_{i}^{2}}\cos\left(2\pi\omega_{2}t\right)-\sin\left(2\pi\omega_{1}t\right)\cos\left(4\pi s_{i}\right)\\ &&+\sin\left(2\pi\omega_{2}t\right)\sin\left(\pi s_{i}\right)\\
\delta_{t}^{\left(2\right)}&:=&e^{s_{i}^{2}}\sin\left(2\pi\omega_{1}t\right)+\cos\left(2\pi\omega_{1}t\right)\cos\left(4\pi s_{i}\right),
\end{eqnarray*}
\noindent where $\omega_{1}=\{0.1, 0.5\}$, $\omega_{2}=\{0,0.25\}$, and $X_{t}^{\left(j\right)}$ are error terms for $j=1,2$. The error terms follow four models drawn directly from \cite{haghbin2019}, one being a Gaussian white noise and the other three coming from a functional autoregressive model of order 1 (FAR1) given by
\begin{equation}\label{FAR1}
X_{t}\left(s\right):=\Psi X_{t-1}\left(s\right)+\epsilon_{t}\left(s\right),\nonumber
\end{equation}
\noindent where the collection $\{\epsilon_{t}\left(s\right)\}_{t=1}^{N}$ are taken as independent functions of Brownian motion over the unit interval and $\Psi$ is an integral operator with kernel
\begin{equation}\label{psi}
\psi\left(s,u\right):=\gamma_{0}\left(2-(2s-1)^{2}-(2u-1)^{2}\right). \nonumber
\end{equation}
We choose $\gamma_{0}$ such that the norm of $\Psi$, given as
\begin{equation}\label{hibsch norm}
\norm{\Psi}^{2}:=\int_{0}^{1}\int_{0}^{1}\abs{\psi\left(s,u\right)}^{2}dsdu, \nonumber
\end{equation}
\noindent takes on values of $0$, $0.5$, or $0.9$ in order to obtain our autoregressive models. Due to the presence of a trend component and two frequencies, we require five components to reconstruct the true structures which is due to the fact that each of the two frequencies is expressed in a sine and a cosine term. We compare reconstruction results of MFSSA, HMFSSA, FSSA performed on each covariate independently of one another, MSSA (HMSSA), and DFPCA ran on each covariate independently of one another. For MSSA we specify that the data matrix, $Q$, follows the form
\begin{equation}\label{Q mssa sim}
Q:=\left[Q_{1}, Q_{2}\right]^{\top},\nonumber
\end{equation}
\noindent such that $Q_{j}:=\left[Y_{t}^{\left(j\right)}\left(s_{i}\right)\right]_{i=1,\dots,100}^{t=1,\dots,N}$ for $j=1,2$ with $i$ being representative of rows of $Q_{j}$ and $t$ of columns. For all of the SSA-based algorithms we set $L=\{20,40\}$ and for all algorithms, we measure the error of each reconstruction with the following root mean square error (RMSE)
\begin{equation}\label{RMSE}
\text{RMSE}:=\sqrt{\frac{1}{N\times n \times p}\sum_{j=1}^{p}\sum_{t=1}^{N}\sum_{i=1}^{n}\left(y_{t}^{\left(j\right)}\left(s_{i}\right)-\hat{y}_{t}^{\left(j\right)}\left(s_{i}\right)\right)^{2}}, \nonumber
\end{equation}
\noindent where $\hat{y}_{t}^{\left(j\right)}\left(s_{i}\right)$ is the reconstruction of covariate $j$, at time point $t$, evaluated at point $s_{i}$. For every unique combination of parameters and error terms, we repeat $100$ times and report the mean of the RMSE's in the following plots whose vertical axes are taken over a log scale.

\begin{figure}[H]
	\begin{subfigure}[b]{\textwidth}
		\centering
		\includegraphics[page=1,width=.84\textwidth]{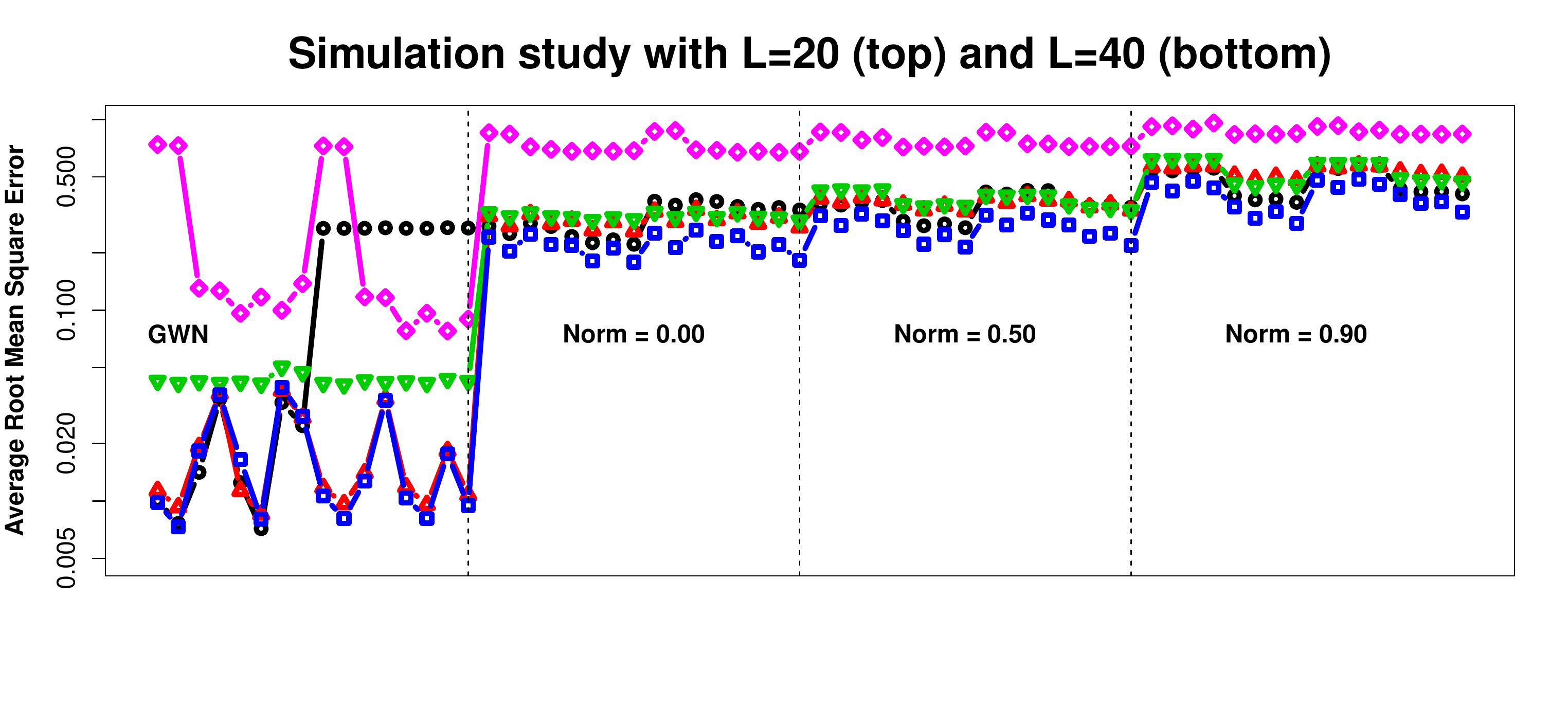}
		\label{fig:sim-A}
	\end{subfigure}	
	
	\begin{subfigure}[b]{\textwidth}
		\vspace{-15mm}
		\centering
		\includegraphics[page=2,width=.84\textwidth]{graphs.pdf}
		\label{fig:sim-B}
	\end{subfigure}
\caption{Simulation Study}
\label{fig:sim_study}
\end{figure}

We see in the top plot that $L=20$ and in the bottom plot, $L=40$, while the vertical lines separate out the simulated data by noise models and in addition, each tick mark on the horizontal should be read as $[N,\omega_{1},\omega_{2},k]$. From these two subfigures, we find that MFSSA almost always outperforms other techniques of dimension reduction for a MFTS while HMFSSA also outperforms other techniques occasionally.

\subsection{Application to Remote Sensing and Weather Station Data}
It is well known that the amount of vegetation present in a region is closely related to the temperature of that same area. Researchers can use this correlation to get a better understanding of how the vegetation and temperature in a region changes over time together through use of multivariate analysis techniques. Data that tracks the intraday hourly mean temperature, in celsius, for a variety of United States weather stations is available for download from \cite{tempdata}. In addition, Satellite images of varying resolutions, regions, time periods, spectral bands, and their variants have been made available for download and analyzed using various techniques \citep{tuck2014}. The normalized difference vegetation index (NDVI) measure, which is bounded between zero and one, is used to track the amount of vegetation, is computed as the difference of the near-infrared and red bands which is then divided by the sum of the same spectral quantities \citep{lambin1999}. NDVI values closer to one are indicative of more vegetation being present while values closer to zero are indicative of less vegetation. It is common practice to average the NDVI measures of each image to form a time series and then analyze it with techniques such as X12-ARIMA \citep{panuju2012}. The issue with this approach is that two different densities that correspond to two different NDVI images might have similar sample means and to this end, more informative approaches should be used. The work of \cite{haghbin2019} estimated a density for each NDVI image taken of a region of Jambi, Indonesia in 16 day increments between February 18, 2002, and July 28, 2019. They then applied FSSA to the time series of densities and discovered a trend component indicating a loss of vegetation over the course of a decade that was not detected by other techniques.

It was determined that using MSSA over SSA can lead to richer analysis of correlated data \citep{golyandina2012}. If a variable with strong seasonality components and another variable with strong mean components are included together in an MSSA analysis, we expect to find strong seasonality and mean component reflected in the singular values and singular vectors. To illustrate this concept continues into the functional realm, we use a bivariate example of intraday hourly mean temperature curves and NDVI images of a parallelogram shaped region just east of Glacier National Park in Montana, U.S.A. located between longitudes of $113.30^{\circ} \text{W} - 113.56^{\circ} \text{W}$ and latitudes of $48.71^{\circ} \text{N} - 48.78^{\circ} \text{N}$  starting January 1, 2008 and ending September 30, 2013 every 16 days. We start by applying FSSA with a lag of $45$ to the functional curves and images separately from one another, where this choice of lag captures annual behavior in the MFTS, and we obtain the following plots of the singular vectors.

\begin{figure}[H]\label{fssatempndvi}
	\begin{subfigure}{.49\linewidth}
		\centering
		\includegraphics[page=1,width=.8\textwidth]{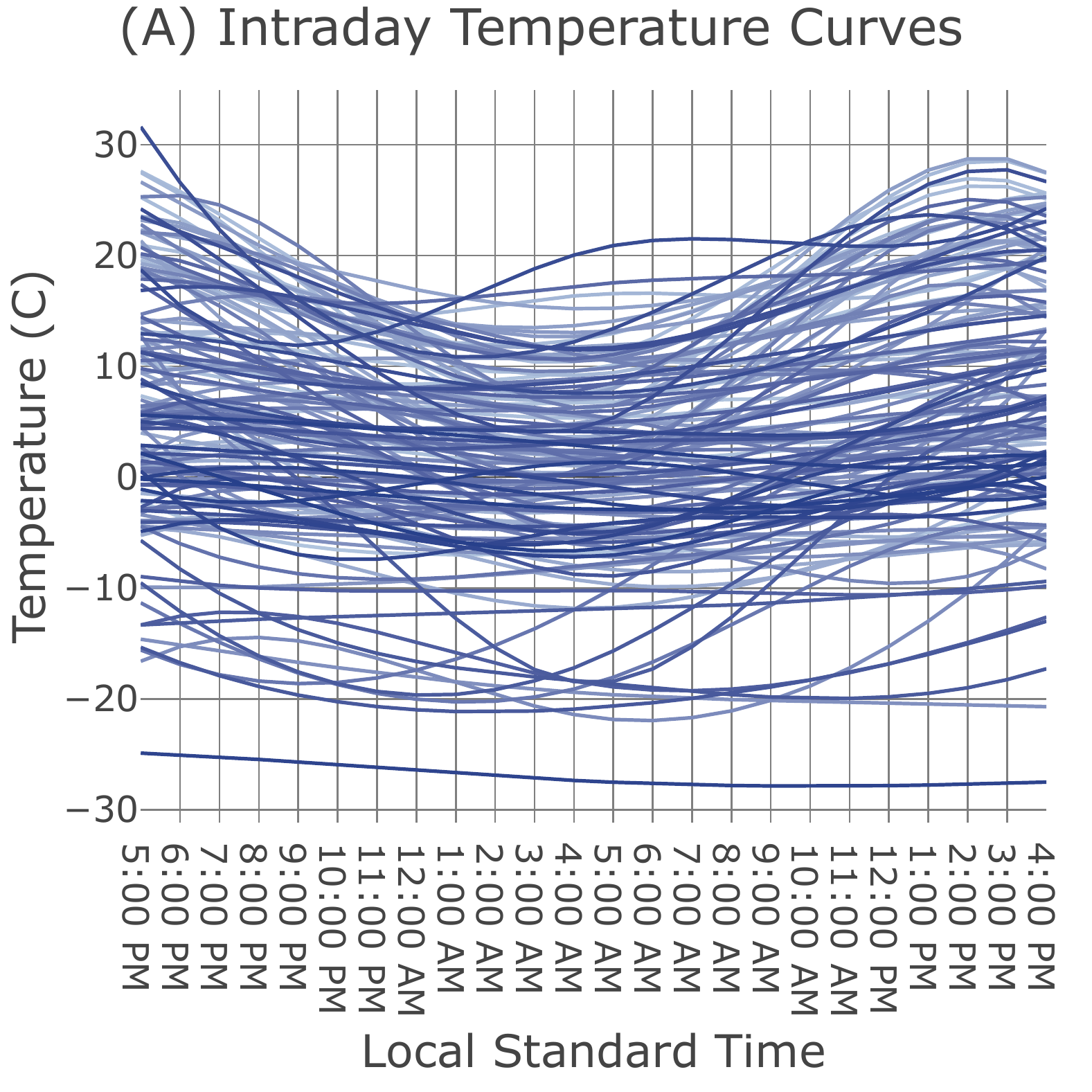}
		\label{fig:temp_A}
	\end{subfigure}	
		\centering
	\begin{subfigure}{.5\linewidth}
		\includegraphics[trim = 0 0 0 10, page=1,width=.8\textwidth]{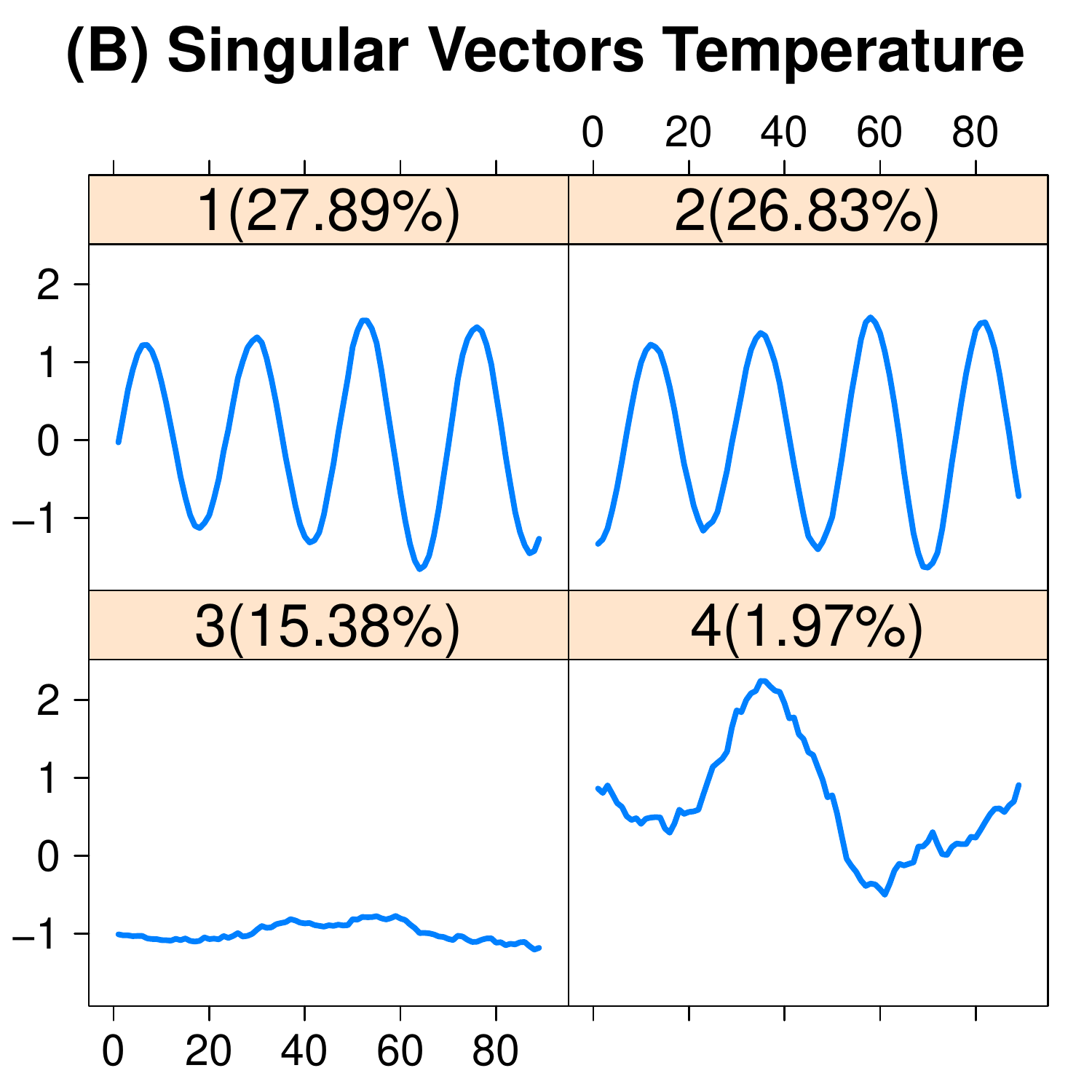}
		\label{fig:temp_vecs_B}
	\end{subfigure}
	\begin{subfigure}{.49\textwidth}
		\centering
		\includegraphics[trim = 0 0 0 0,clip,page=1,width=.8\textwidth]{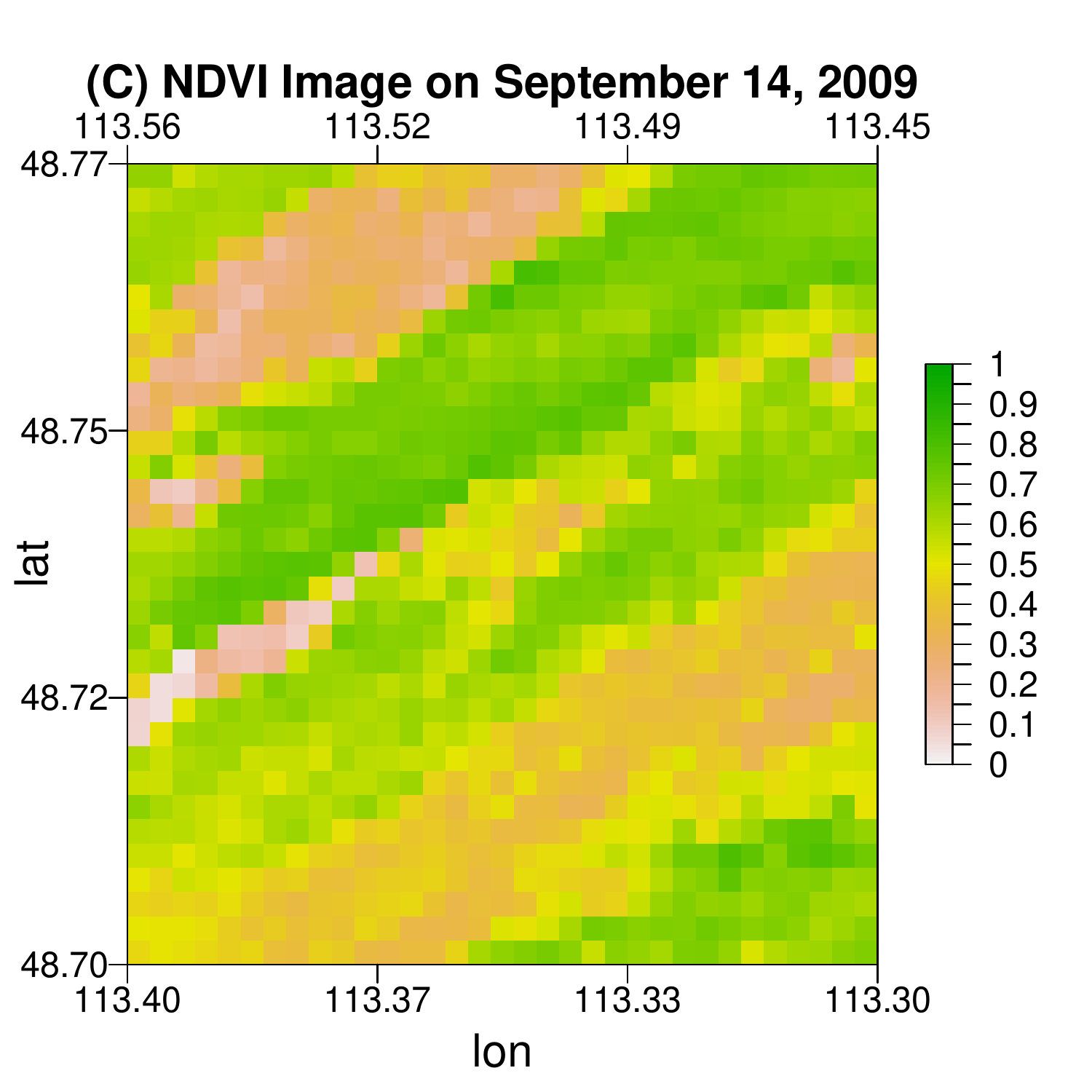}
		\label{fig:NDVI_C}
	\end{subfigure}
	\begin{subfigure}{.5\textwidth}
		\includegraphics[trim= 0 0 0 10,page=1,width=.8\textwidth]{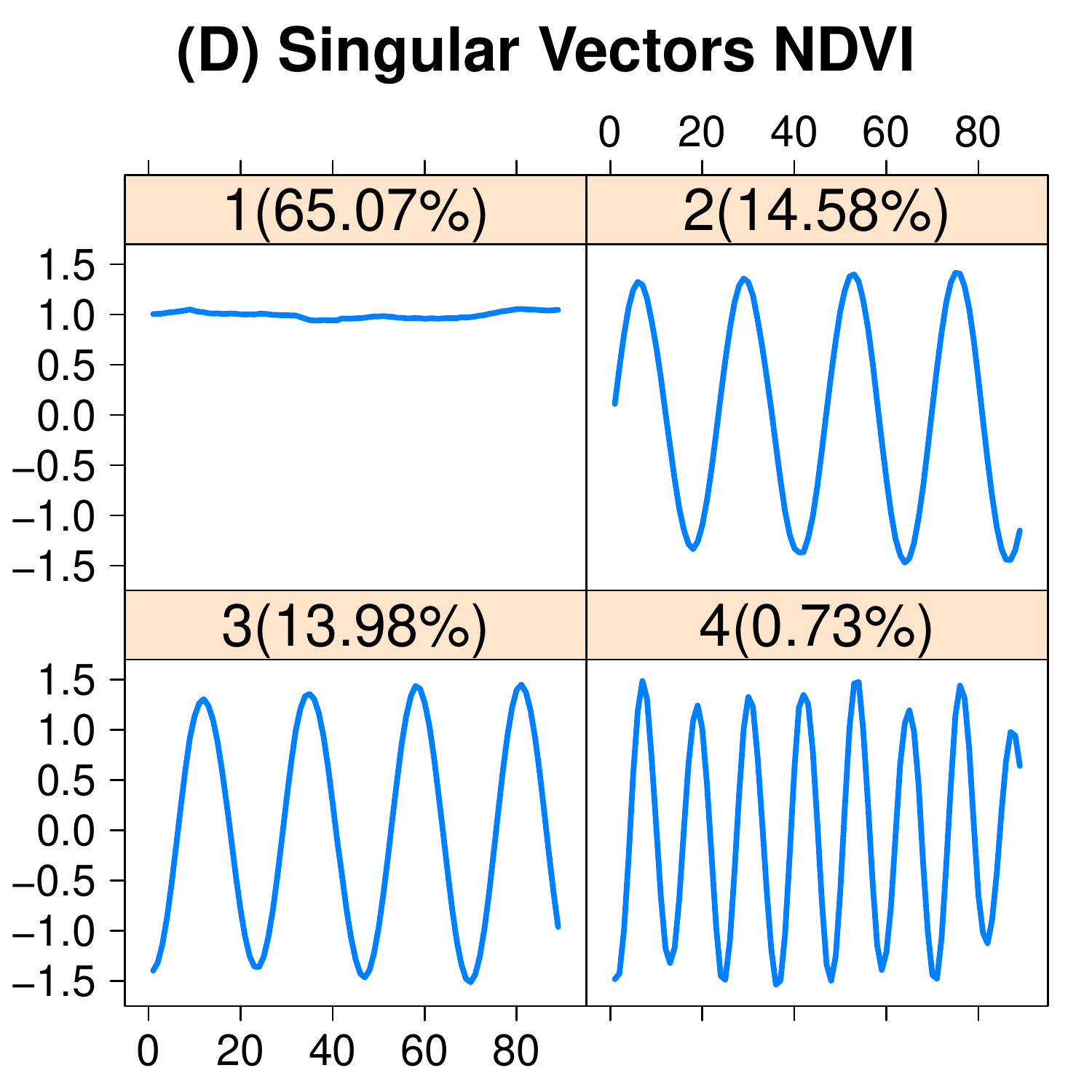}
		\label{fig:NDVI_vecs_D}
	\end{subfigure}
\caption{FSSA on Intraday Temperature Curves and FSSA on NDVI Functional Images}
\label{fig:fssa_temp_ndvi}
\end{figure}

\noindent It is clear from plot (B) of Figure \ref{fig:fssa_temp_ndvi} that there exists a strong seasonality component in the intraday temperature curves of plot (A) accounting for $54.72\%$ of the variation in the data while a mean behavior component accounts for $15.38\%$ of the variation in the data. We also see from plot (D) of Figure \ref{fig:fssa_temp_ndvi} that the mean component captures $65.07\%$ of the variation of the NDVI images data where plot (C) is one such observations while the seasonality components only account for $28.56\%$ of the variation of the data. We normalize the intraday temperature curves by dividing each sampling point by the standard deviation of all the sampling points since the NDVI images have values that are significantly smaller. We now apply MFSSA with a lag of $45$ to the normalized intraday temperature curves and NDVI images in a bivariate analysis to obtain the following plots.

\begin{figure}[H]
	\begin{subfigure}{0.32\textwidth}
		\includegraphics[trim=0 0 0 10,width=.98\textwidth]{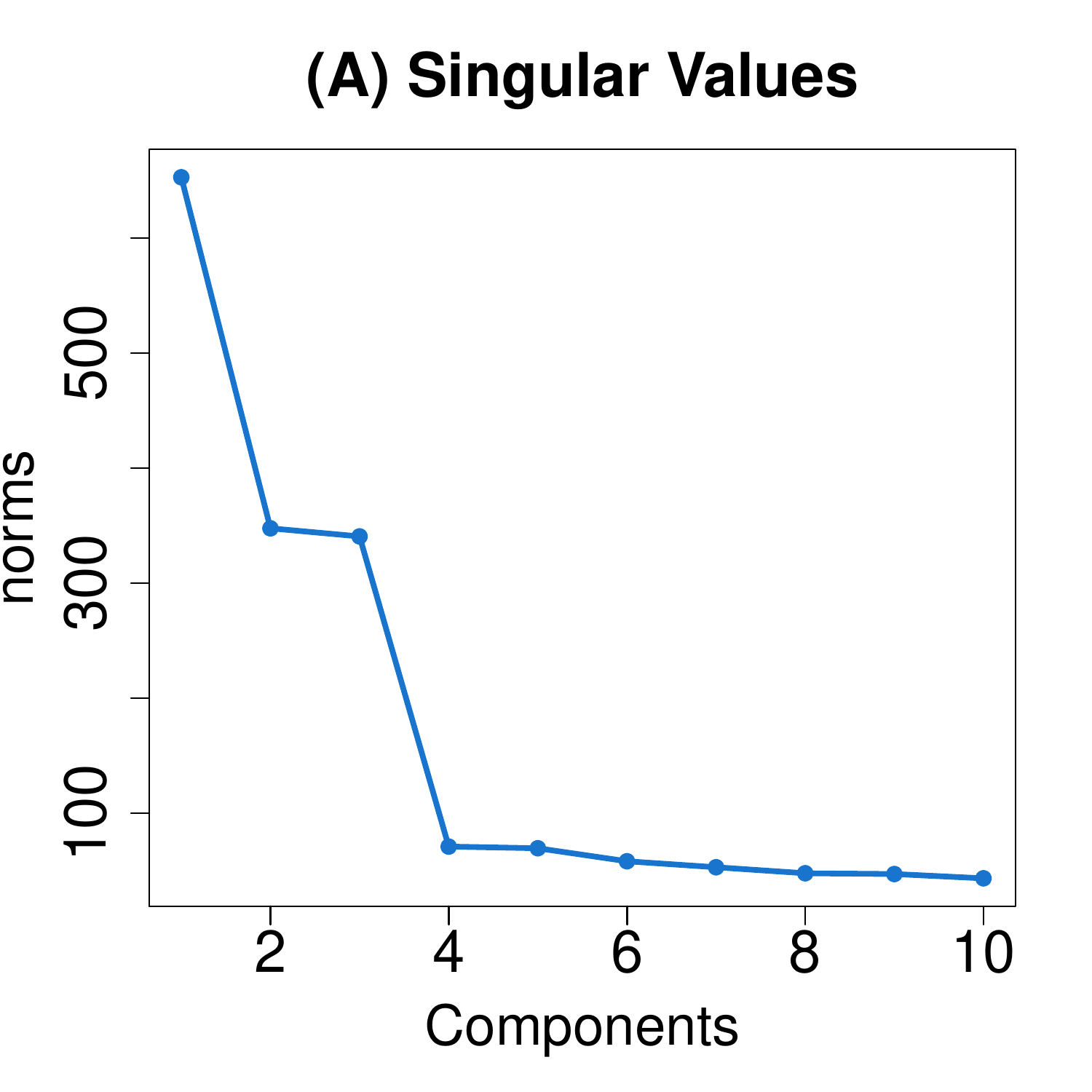}
		\label{fig:mfssa_sing_vals_A}
	\end{subfigure}	
	\begin{subfigure}{0.33\textwidth}
		\includegraphics[trim= 0 0 0 0,page=1,width=.98\textwidth]{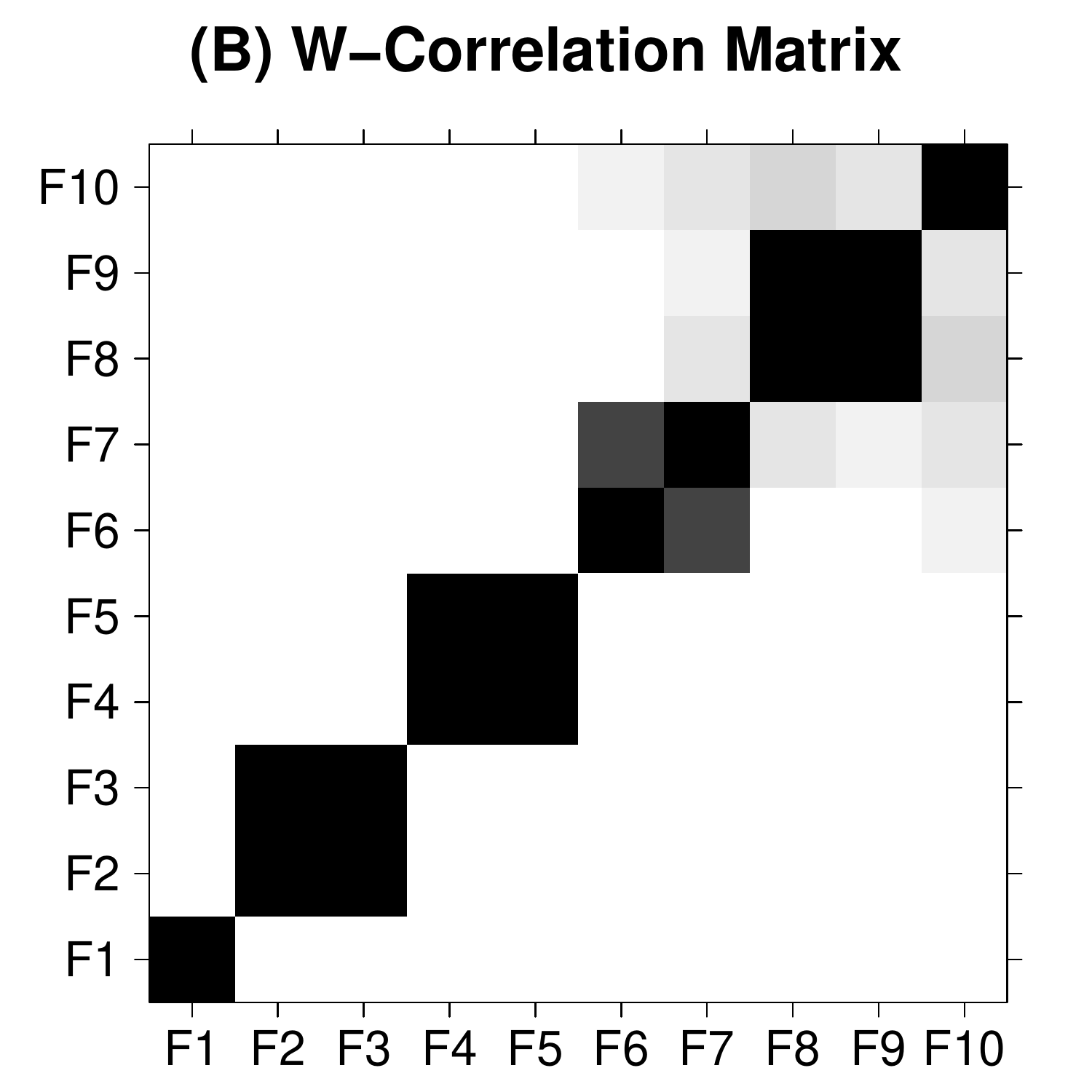}
		\label{fig:mfssa_wcor_B}
	\end{subfigure}
	\begin{subfigure}{0.33\textwidth}
		\includegraphics[trim=0 0 0 0,page=1,width=.98\textwidth]{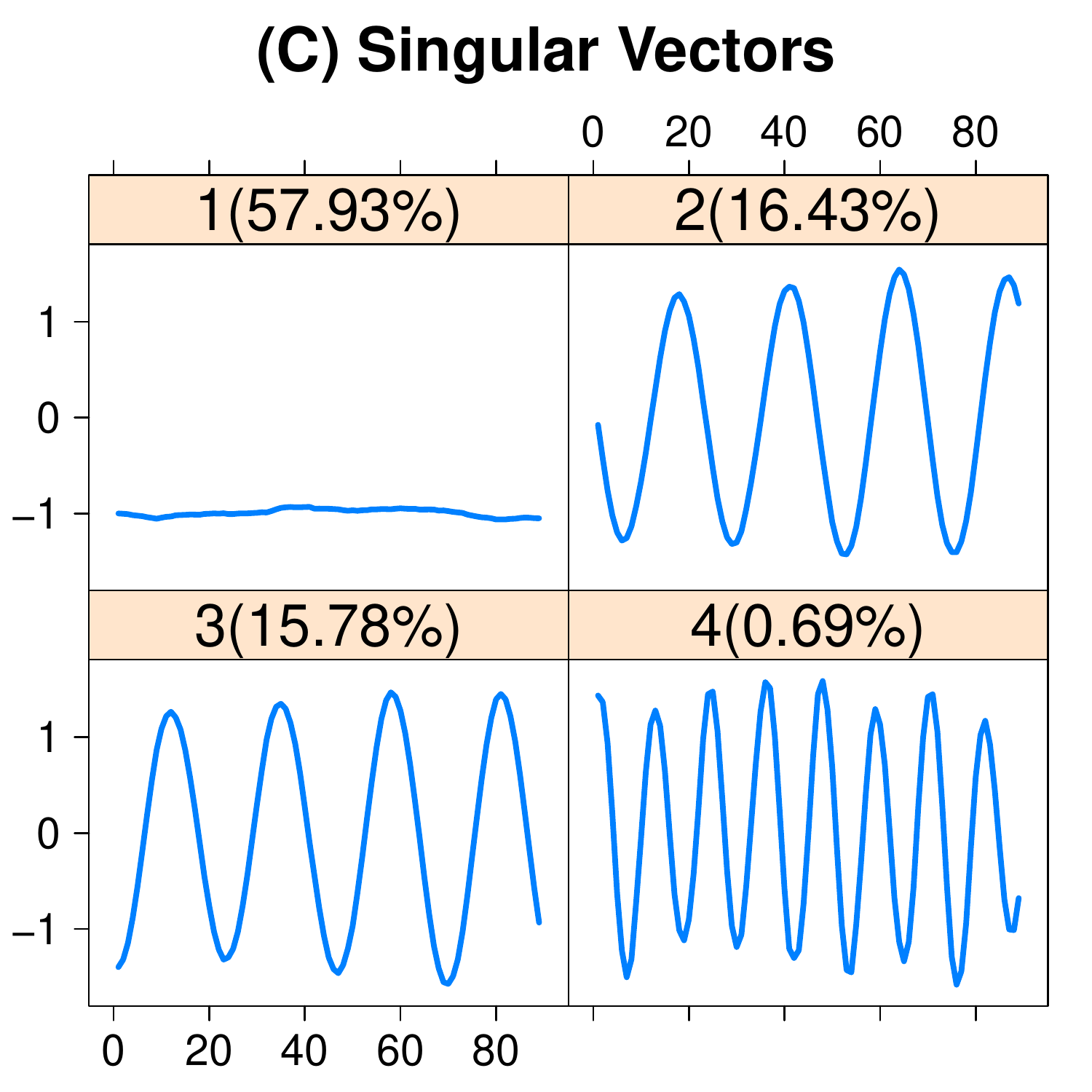}
		\label{fig:mfssa_sing_vecs}
	\end{subfigure}
\caption{MFSSA Exploratory Plots}
\label{fig:mfssa_temp_NDVI}
\end{figure}

Plots (A) and (B) of Figure \ref{fig:mfssa_temp_NDVI} show that component one should be grouped by itself, two should be grouped with three, and four with five. Plot (C) of Figure \ref{fig:mfssa_temp_NDVI} shows that in the bivariate analysis, the mean component becomes dominant with the seasonal components taking on the second and third main sources of variation. This shows that combining the temperature curves and NDVI images functional data into a bivariate analysis reveals a stronger mean component as opposed to the weaker mean component seen in plot (E) of Figure \ref{fig:fssa_temp_ndvi}.

\section{Discussion}

Throughout this paper, we presented MFSSA as a novel technique of dimension reduction of a MFTS. We found that the MFSSA problem is solved by performing VMFSSA and we also developed HMFSSA, presented in supplementary material, as another approach but found that it was more restrictive and not as informative as MFSSA. We also developed MFSSA to be able to handle functions taken over different dimensional domains to uncover a more dominant mean component for the intraday temperature curves/NDVI images bivariate analysis. The MFSSA algorithm is available for use in the \pkg{Rfssa} package \citep{rfssapackage}, available through CRAN.
\bibliographystyle{apa}
\bibliography{Mybib}


\newpage
\setcounter{page}{1}
\renewcommand{\thefigure}{S\arabic{figure}}
\renewcommand{\thetable}{S\arabic{table}}
\renewcommand{\theequation}{S\arabic{equation}}
\renewcommand{\thesection}{S\arabic{section}}
\setcounter{figure}{0}   
\setcounter{table}{0}   
\setcounter{equation}{0}
\setcounter{section}{0}   
\section*{Supplementary Materials}
The supplementary material includes plots and animations of the left singular functions of our real data study in the manuscript, another remote sensing real data study example, and the full development of HMFSSA. We also include proofs of the lemmas and propositions of the manuscript.
\section{Left Singular Functions of MFSSA}

In this section, we build on the real data study presented in the manuscript by presenting the left singular functions. As mentioned, we apply FSSA to the temperature curves and NDVI images separately, We also implement MFSSA to the temperature curves and NDVI images together, both with a lag of $45$, to obtain the following.

\begin{figure}[H]\label{fssatempfuns}
	\begin{subfigure}{.49\linewidth}
		\centering
		\includegraphics[page=1,width=.85\textwidth]{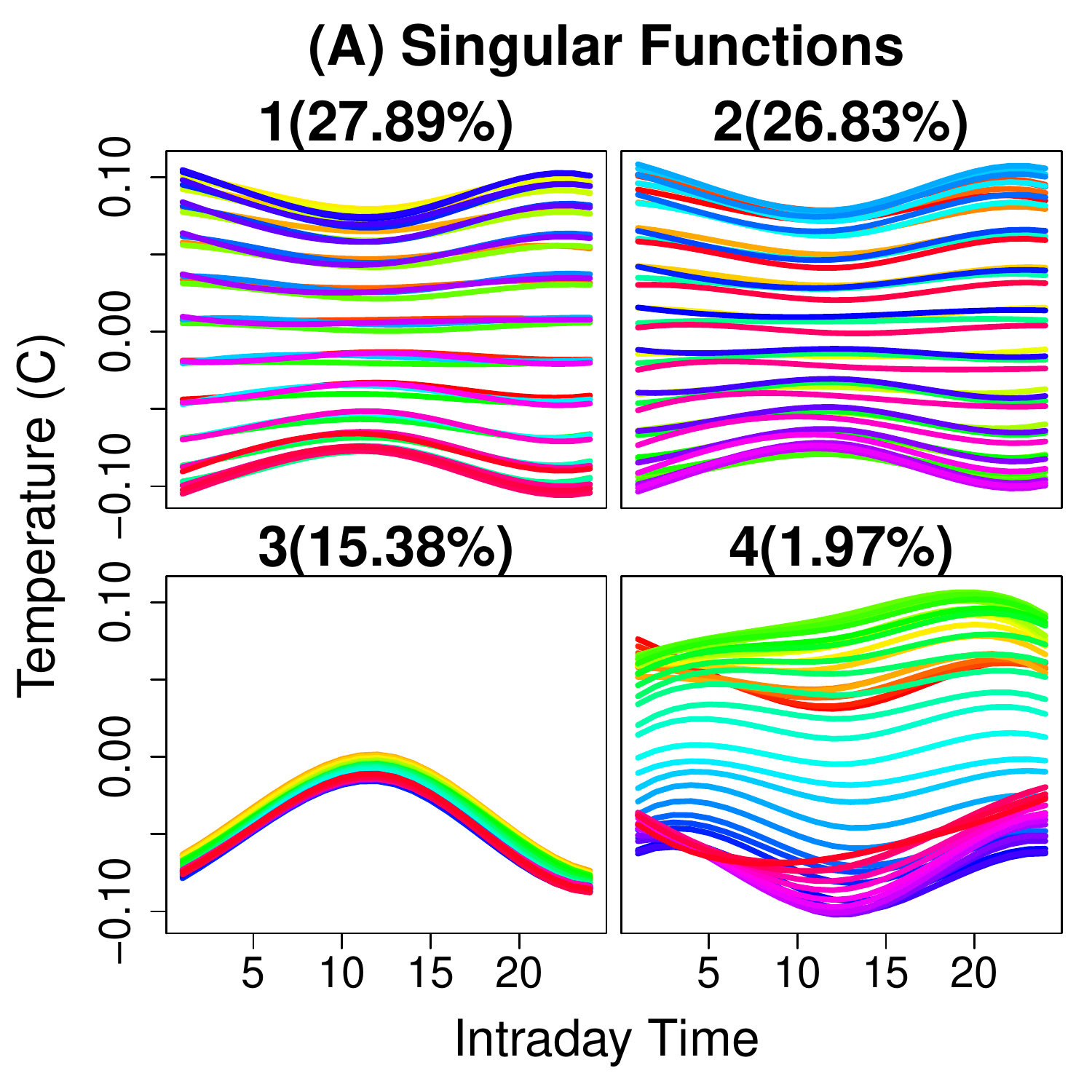}
		\label{fig:curves_A}
	\end{subfigure} 
	\begin{subfigure}{0.50\linewidth}
		\centering
		\includegraphics[page=1,width=.85\textwidth]{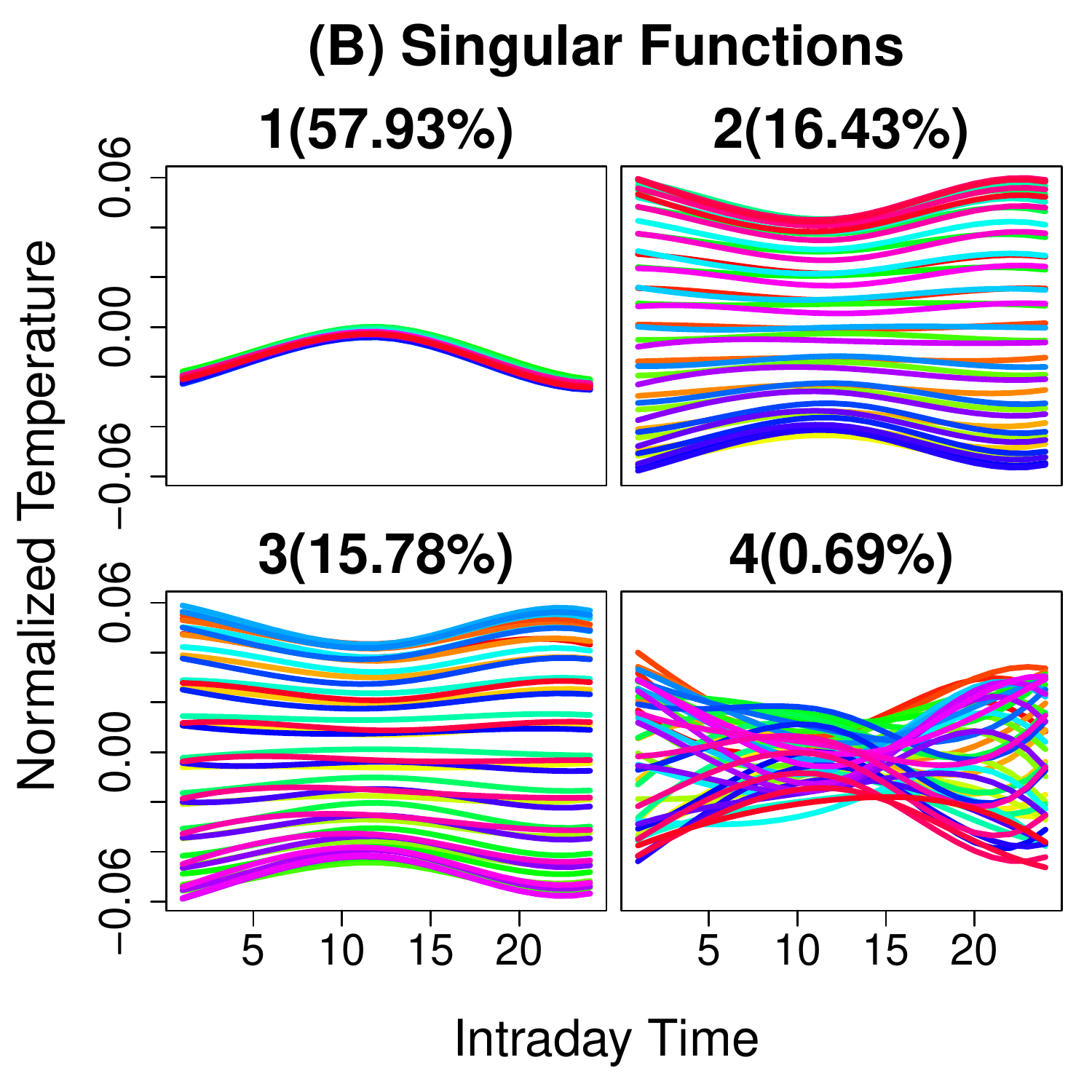}
		\label{fig:mfssa_curves_B}
	\end{subfigure}
	\begin{subfigure}{.49\linewidth}
		\centering
		 \animategraphics[controls,width=.85\textwidth]{.80}{temp_funs_ani}{0}{44}
 		\label{fig:ani_curves_C}
	\end{subfigure} 
	\begin{subfigure}{0.50\linewidth}
		\centering
		 \animategraphics[controls,width=.85\textwidth]{.80}{mfssa_temp_funs_ani}{0}{44}
 		\label{fig:mfssa_ani_curves_D}
	\end{subfigure}
\caption{Subfigures (A), (C): FSSA Left Singular Functions. Subfigures (B), (D): MFSSA Left Singular Functions}
\label{fig:fssa_funs_temp}
\end{figure}

\noindent Plot (A) of Figure \ref{fig:fssa_funs_temp} shows all $L=45$ functions of the first four left singular functions of FSSA for the temperature data while plot (C) steps through each function in an animation. Plot (B) of Figure \ref{fig:fssa_funs_temp} shows all $L=45$ functions of the first four left singular functions of MFSSA for the temperature data while plot (D) steps through each function in an animation. We see in the temperature data, that when MFSSA is applied, the mean component becomes stronger. We apply FSSA to the images with a lag of $45$ and compare the resulting left singular functions for the NDVI images to those we obtain via the MFSSA analysis in the following animations. 

\begin{figure}[H]\label{mfssatempNDVIfuns}
	\begin{subfigure}{.49\linewidth}
		\centering
		 \animategraphics[controls,width=.9\textwidth]{.80}{NDVI_fun_ani}{0}{44}
 		\label{fig:images_B}
	\end{subfigure} 
	\begin{subfigure}{0.50\linewidth}
		\centering
		 \animategraphics[controls,width=.9\textwidth]{.80}{mfssa_NDVI_fun_ani}{0}{44}
 		\label{fig:images_B}
	\end{subfigure}
\caption{Subfigure (A): FSSA NDVI Images. Subfigure (B): MFSSA NDVI Images}
\label{fig:mfssa_funs_temp_NDVI}
\end{figure}

\noindent Here, we see little difference between the animations.
 
\section{MFSSA Applied to Remote Sensing Density Curves}

To further show that MFSSA enriches data analysis of correlated variables, we use a bivariate example of near-infrared (NIR) and shortwave infrared (SWIR) images taken every eight days of a region just outside of the city of Jambi, Indonesia between $103.61^{\circ} \text{E} -  103.68^{\circ} \text{E}$ and $1.67^{\circ} \text{S} -  1.60^{\circ} \text{S}$ over the timeline of February 18, 2000 and November 25, 2019. The wavelength of the NIR images range from 841-876 nanometers (nm) and the wavelength of the SWIR images are within the values of 2105-2155 nm. NIR light can be used for imaging vegetation as it is used in the calculation of the NDVI measure \citep{lambin1999} while shortwave infrared is often used for imaging the moisture content in soil where a lower surface reflectance (SR) corresponds to higher moisture content \citep{shin2017}. As mentioned in \cite{prasetyo2016}, it appears that this particular part of the Jambi province was a hot spot for controlled fires between 2001 and 2015 and this loss of vegetation over the course of about a decade will be reflected in lower NIR and higher SWIR SR values as time moves on. We obtain the KDEs of both the NIR and the SWIR SR images using Silverman's rule of thumb \citep{silverman1986} which we then project onto a cubic B-spline basis where the degrees of freedom are chosen using the GCV criterion. In addition, we replaced outliers in the SWIR densities with the average of densities from the preceding and proceeding days. Similar results, as compared to the following, still hold even if the outliers are not removed. Applying FSSA with a lag of $45$ to the NIR and SWIR densities separately, where this choice of lag approximately captures annual behavior, gives the following exploratory plots.

\begin{figure}[H]\label{fssanirswir}
	\begin{subfigure}{.49\linewidth}
		\centering
		\includegraphics[page=1,width=.8\textwidth]{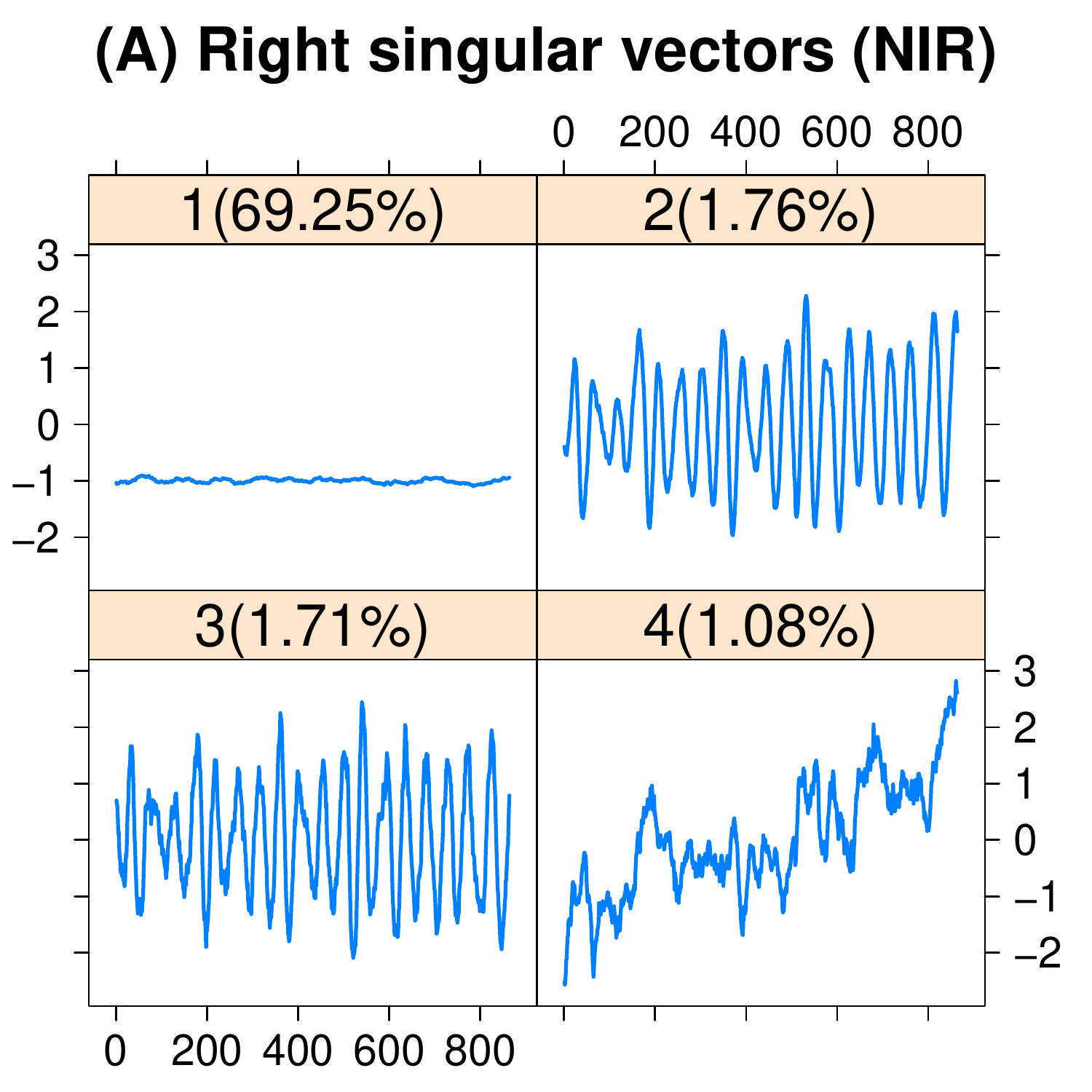}
		\label{fig:NDVI-A}
	\end{subfigure}	
		\centering
	\begin{subfigure}{.5\linewidth}
		\includegraphics[trim = 0 0 0 10, page=2,width=.8\textwidth]{swir.pdf}
		\label{fig:NDVI-B}
	\end{subfigure}
	\begin{subfigure}{.49\textwidth}
		\centering
		\includegraphics[trim = 0 0 0 0,clip,page=3,width=.8\textwidth]{swir.pdf}
		\label{fig:NDVI-C}
	\end{subfigure}
	\begin{subfigure}{.5\textwidth}
		\includegraphics[trim= 0 0 0 10,page=4,width=.8\textwidth]{swir.pdf}
		\label{fig:NDVI-D}
	\end{subfigure}
\caption{FSSA on NIR and SWIR Densities}
\label{fig:swi_fssa}
\end{figure}

Figure \ref{fig:swi_fssa} subfigures (A) and (B) give us the right singular vectors and left singular functions of the NIR densities while Figure \ref{fig:swi_fssa} subfigures (C) and (D) are the right singular vectors and left singular functions of the SWIR densities. We find that applying FSSA to the NIR densities captures seasonality in the second and third components while trend is present in the fourth component similar to the NDVI results of \cite{haghbin2019}. Applying FSSA to the SWIR densities shows that trend is a more dominant behavior captured in the second component as compared to the seasonal behaviors captured in components three and four. Applying MFSSA decomposition with a lag of $45$ to the bivariate NIR/SWIR example, where this lag is chosen to capture annual behavior, gives the following exploratory plots.

\begin{figure}[H]
	\begin{subfigure}[b]{0.32\textwidth}
		\includegraphics[trim=0 0 0 10,width=.98\textwidth]{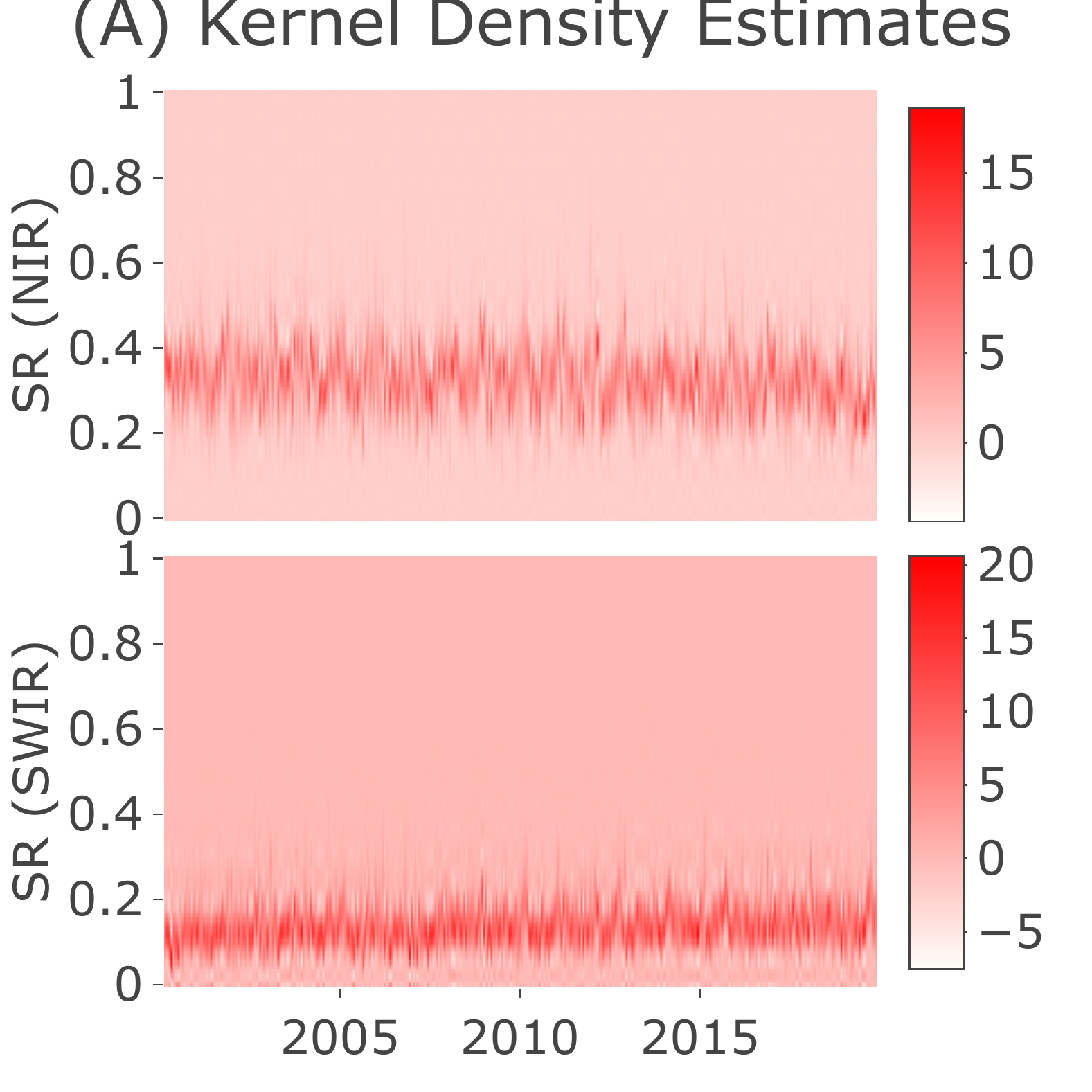}
		\label{fig:swir_nr-A}
	\end{subfigure}	
	\begin{subfigure}[b]{0.33\textwidth}
		\includegraphics[trim= 0 0 0 0,page=1,width=.98\textwidth]{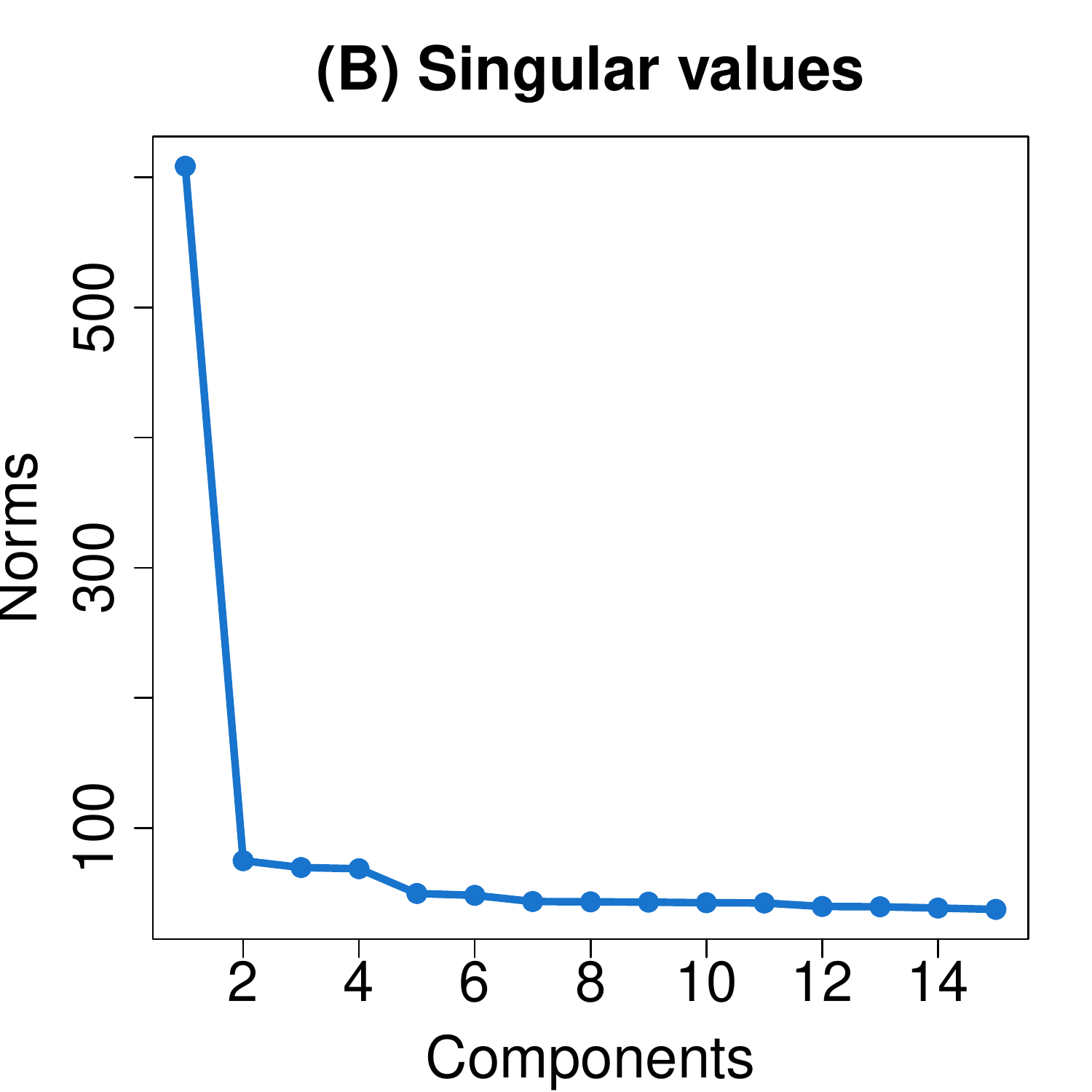}
		\label{fig:swir_nr-B}
	\end{subfigure}
	\begin{subfigure}[b]{0.33\textwidth}
		\includegraphics[trim=0 0 0 0,page=2,width=.98\textwidth]{swir_nr.pdf}
		\label{fig:swir_nr-C}
	\end{subfigure}
	\begin{subfigure}[b]{0.33\textwidth}
		\includegraphics[trim=0 25 0 10,page=3,width=.98\textwidth]{swir_nr.pdf}
		\label{fig:swir_nr-D}
	\end{subfigure}
	\begin{subfigure}[b]{0.33\textwidth}
		\includegraphics[trim=0 15 0 10,page=4,width=.93\textwidth]{swir_nr.pdf}
		\label{fig:swir_nr-E}
	\end{subfigure}
	\begin{subfigure}[b]{0.32\textwidth}
		\includegraphics[trim=0 10 0 0,page=5,width=.93\textwidth]{swir_nr.pdf}
		\label{fig:swir_nr-F}
	\end{subfigure}
\caption{KDEs of NIR and SWIR Images as well as MFSSA Exploratory Plots}
\label{fig:mfssa_nir_swir}
\end{figure}

The bivariate FTS can be found in Figure \ref{fig:mfssa_nir_swir} subfigure (A) while Figure \ref{fig:mfssa_nir_swir} subfigures (B) and (C) are plots of singular values and w-correlation respectively. See that Figure \ref{fig:mfssa_nir_swir} subfigure (D) gives us our MFSSA right singular vectors which showcases the weights that are multiplied by the left singular functions shown in Figure \ref{fig:mfssa_nir_swir} subfigures (E) and (F). Since we are performing MFSSA, we obtain $45$ left eigenfunctions that correspond to the NIR densities as well as another set of $45$ left eigenfunctions that correspond to the SWIR densities. Notice the trend behavior for the NIR densities is present in component two as according to Figure \ref{fig:mfssa_nir_swir} subfigure (E) which indicates that adding SWIR densities into the analysis with the NIR densities created a more pronounced trend result as compared with Figure \ref{fig:swi_fssa} subfigure (B). To this end, we find that performing a bivariate analysis on the NIR/SWIR densities enriched our data analysis as expected.

\section{HMFSSA}

We begin this section with our discussion of moving from HMSSA to HMFSSA. As we clarified in subsection 5.1 of the manuscript, we need to assume $T:=T_{1}=\cdots=T_{p}$, $\mathbb{F}:=L^2(T)$, and $\pmb{x}_k^{(j)}$'s belong to a common space $\mathbb{F}^L$, for $j=1,\cdots,p$. Notice that while the domain for each variable is the same, one may evaluate each variable at different points along $T$. We present the four main steps of the HMFSSA algorithm in the following subsection.

\subsection{Embedding, Decomposition, Grouping, and Reconstruction}\label{embdechmfssa}
We choose $0 < L < \frac{N}{2}$, set $K=N-L+1$, and we define the linear operator $\utilde{\mathcal{X}}:\mathbb{R}^{pK} \rightarrow \mathbb{F}^{L}$ given by
\begin{equation}
\utilde{\mathcal{X}}\left(\pmb{a}\right)=\sum_{j=1}^{p}\sum_{k=1}^{K}a_{k}^{\left(j\right)}\pmb{x}_k^{(j)}, \qquad \pmb{a}\in\mathbb{R}^{pK} \nonumber
\end{equation}
\noindent which follows a similar form as compared to equation (5.1) of the manuscript. The operator, $\utilde{\mathcal{X}}$, is block Hankel, has rank $1\leq \tilde{r} \leq pK$, and we have that $\text{R}\left(\utilde{\mathcal{X}}\right) =\text{sp}\{\pmb{x}_k^{(j)}\}_{k=1,\dots,K}^{j=1,\dots,p}$. It is easy to see from the range of $\utilde{\mathcal{X}}$ why all variables must share a common domain $T$. 

Since $\utilde{\mathcal{X}}$ is a finite rank operator and thus compact, we utilize Theorem 7.6 from \cite{weidmann1980} to obtain the following fSVD for HMFSSA
\begin{equation}\label{hmfssa:decomp}
\utilde{\mathcal{X}}\left(\pmb{a}\right)=\sum_{i=1}^{\tilde{r}}\utilde{\sigma}_{i}\innp{\utilde{\mathbf{v}}_{i}}{\pmb{a}}_{\mathbb{R}^{pK}}\utilde{\boldsymbol{\psi}}_{i}=\sum_{i=1}^{\tilde{r}}\utilde{\sigma}_{i}\utilde{\mathbf{v}}_{i}\otimes\utilde{\boldsymbol{\psi}}_{i}\left(\pmb{a}\right)=\sum_{i=1}^{\tilde{r}}\utilde{\mathcal{X}}_{i}\left(\pmb{a}\right), \nonumber
\end{equation}
\noindent where $\{\utilde{\sigma}_{i}\}_{i=1}^{\tilde{r}}$ are the singular values, $\{\utilde{\mathbf{v}}_{i}\}_{i=1}^{\tilde{r}}$ are the orthonormal right singular vectors that span $\mathbb{R}^{\tilde{r}}$, and $\{\utilde{\boldsymbol{\psi}}_{i}\}_{i=1}^{\tilde{r}}$ are the orthonormal left singular functions that span an ${\tilde{r}}$-dimensional subspace of $\mathbb{F}^{L}$. Also notice that $\{\utilde{\mathcal{X}}_{i}\}_{i=1}^{\tilde{r}}$ are rank one elementary operators similar to those seen in equation (3.4) of the manuscript.

The grouping stage of HMFSSA is similar to the grouping stage of other types of SSA where we form operators $\utilde{\mathcal{X}}_{I_{q}}:\mathbb{R}^{pK} \rightarrow \mathbb{F}^{L}$ for $1\leq q \leq m$. We finish by projecting each $\utilde{\mathcal{X}}_{I_{q}}$ onto the subspace of block Hankel operators that map from $\mathbb{R}^{pK}$ to $\mathbb{F}^{L}$ to form a collection of $q$ reconstructed MFTS where the projection is completed blockwise using the diagonal averaging technique of \cite{haghbin2019}.

\subsection{HMFSSA Implementation}
Implementation of HMFSSA is similar to that of \cite{haghbin2019} since $\utilde{\mathcal{X}}$ maps to $\mathbb{F}^{L}$. Let $\{\nu_{k}\}_{k \in \mathbb{N}}$ be a known basis of the space $\mathbb{F}$ such that any $x \in \mathbb{F}$ can be projected onto the subspace $\mathbb{H}_d:=\text{sp}\{\nu_{i}\}_{i=1}^{d}$. As such, each $\hat{x} \in \mathbb{H}_d$ can be represented as
\begin{equation}\label{hmfssaimpexp}
\hat{x}=\sum_{k=1}^{d}c_{k,i}\nu_{k}, \quad i=1,\dots,N, \quad c_{k,i} \in \mathbb{R}, \nonumber
\end{equation}
\noindent Let $\mathbb{H}_{d}^{L}$ be the $d$-dimensional subspace formed from the Cartesian product of $L$ copies of $\mathbb{H}_d$, then the rest of the work in defining basis elements of $\mathbb{H}_{d}^{L}$ follows directly from \cite{haghbin2019}. The work involving the expansion of the lagged vectors, the range of $\utilde{\mathcal{X}}$, the definition of the coefficient matrix $\mathbf{B}:=\left[b_{k,i}\right]_{i=1,\dots,Ld}^{k=1,\dots,pK}=\left[\mathbf{b}_{1}, \mathbf{b}_{2}, \dots, \mathbf{b}_{pK}\right]_{Ld \times pK}$, and the HMFSSA version of Theorem 4.1  seen in the manuscript, also follows from \cite{haghbin2019} except for the fact that we replace $K$ with $pK$.

\subsection{HMFSSA SWIR/NIR Study}
To show that HMFSSA separates out MFTS behavior based on the covariate, we apply HMFSSA with a lag of $45$ to the NIR/SWIR example and obtain the following plots.

\begin{figure}[H]
	\begin{subfigure}[b]{0.33\textwidth}
		\includegraphics[trim= 0 0 0 0,page=1,width=.98\textwidth]{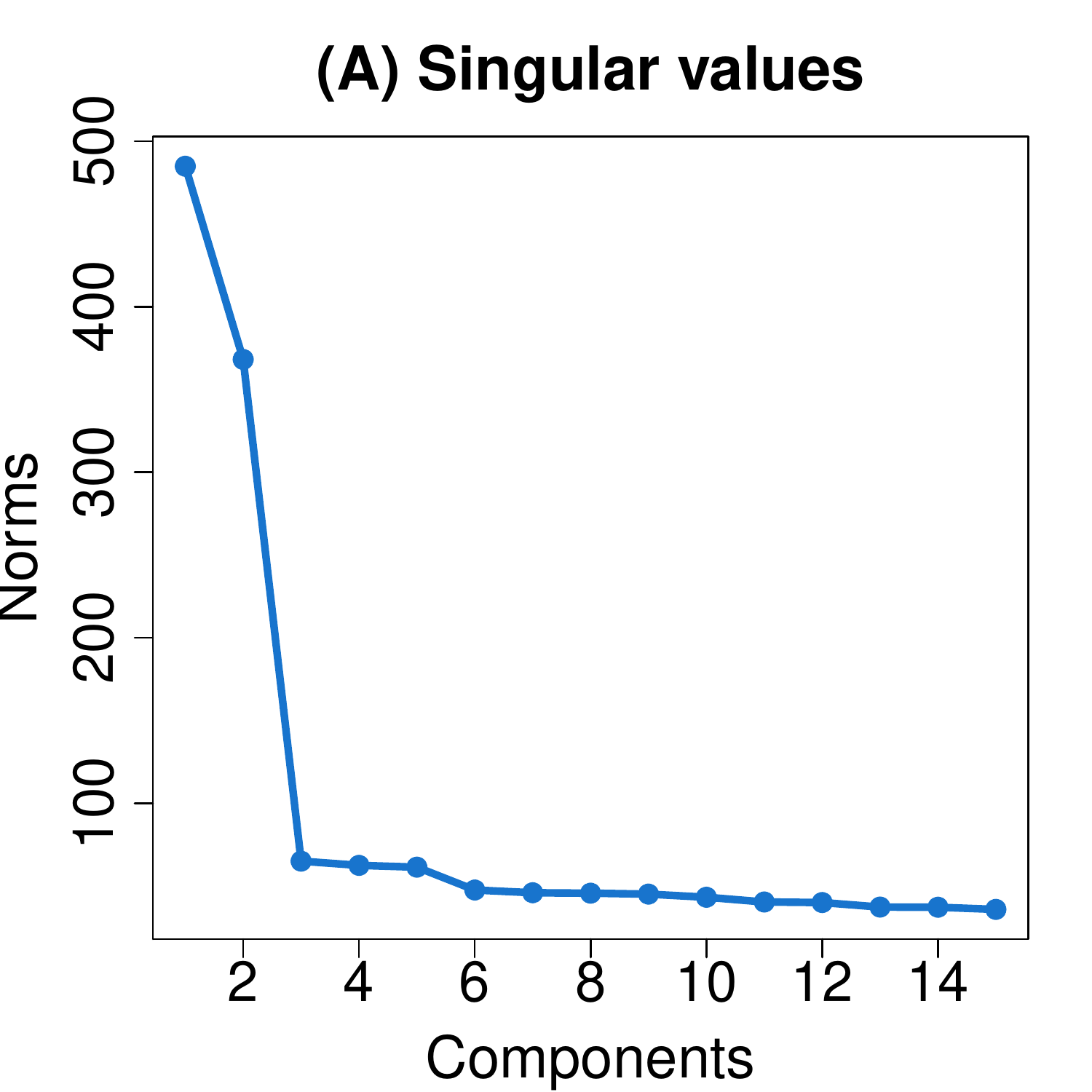}
		\label{fig:swir_nr-A}
	\end{subfigure}	
	\begin{subfigure}[b]{0.33\textwidth}
		\includegraphics[trim= 0 0 0 0,page=2,width=.98\textwidth]{swir_nr_hmfssa.pdf}
		\label{fig:swir_nr-B}
	\end{subfigure}
	\begin{subfigure}[b]{0.32\textwidth}
		\includegraphics[trim= 0 0 0 0,page=3,width=.98\textwidth]{swir_nr_hmfssa.pdf}
		\label{fig:swir_nr-C}
	\end{subfigure}
	\begin{subfigure}[b]{0.32\textwidth}
		\includegraphics[trim= 0 0 0 0,page=4,width=.98\textwidth]{swir_nr_hmfssa.pdf}
		\label{fig:swir_nr-D}
	\end{subfigure}
	\begin{subfigure}[b]{0.33\textwidth}
		\includegraphics[trim= 0 0 0 0,page=5,width=.98\textwidth]{swir_nr_hmfssa.pdf}
		\label{fig:swir_nr-E}
	\end{subfigure}
	\begin{subfigure}[b]{0.33\textwidth}
		\includegraphics[trim= 0 0 0 0,page=6,width=.98\textwidth]{swir_nr_hmfssa.pdf}
		\label{fig:swir_nr-F}
	\end{subfigure}
\caption{HMFSSA Exploratory Plots}
\label{fig:hmfssa_nir_swir}
\end{figure}

In this case, we have $K$ right singular vectors that correspond to NIR densities and $K$ right singular vectors that correspond to SWIR densities. It appears that the first component captures mean behavior of SWIR densities while the second component captures mean behavior of the NIR densities seen in Figure \ref{fig:hmfssa_nir_swir} subfigures (E) and (F) which is confirmed when we compare with Figure \ref{fig:swi_fssa} subfigures (A) and (C). Rather than combining information to create a more pronounced mean component, HMFSSA works to separate out these behaviors by variable which is expected due to the similarity between HMFSSA and FSSA.

\section{Proofs}

\begin{proof}[Proof of Prop. 3.1]

\noindent Notice that since $\text{R}\left(\mathbfcal{X}\right)=\text{sp}\{\pmb{x}_{k}\}_{k=1}^{K}$, then $\mathbfcal{X}$ is a rank $1\leq r \leq K$ operator and thus compact. As such, we have that $\mathbfcal{X}$ is bounded. Let $\pmb{a},\pmb{b} \in \mathbb{R}^{K}$ and $c \in \mathbb{R}$, then we have that
\begin{equation}
\mathbfcal{X}\left(\pmb{a}+c\pmb{b}\right)=\sum_{k=1}^{K}\left(a_{k}+cb_{k}\right)\pmb{x}_{k}=\sum_{k=1}^{K}a_{k}\pmb{x}_{k}+c\sum_{k=1}^{K}b_{k}\pmb{x}_{k}=\mathbfcal{X}\pmb{a}+c\mathbfcal{X}\pmb{b} \nonumber
\end{equation}
\noindent which implies that $\mathbfcal{X}$ is a linear operator. Now let $\mathbf{z} \in \mathbb{H}^{L}$, then
\begin{equation}
\innp{\mathbfcal{X}\pmb{a}}{\mathbf{z}}_{\mathbb{H}^{L}}=\sum_{k=1}^{K}a_{k}\innp{\pmb{x}_{k}}{\mathbf{z}}_{\mathbb{H}^{L}}=\pmb{a}^{\top}\mathbfcal{X}^{*}\mathbf{z}=\innp{\pmb{a}}{\mathbfcal{X}^{*}\mathbf{z}}_{\mathbb{R}^{K}},\quad \pmb{a}\in\mathbb{R}^{K} \nonumber
\end{equation}
\noindent and we have that $\mathbfcal{X}^{*}$ is the adjoint of $\mathbfcal{X}$.
\end{proof}

\begin{proof}[Proof of Prop. 3.2]

\noindent Let $\mathbf{S}=\mathbfcal{X}^{*}\mathbfcal{X}$ be the $K \times K$ variance/covariance matrix for the $L$-lagged vectors of $\mathbfcal{X}$. Since $\mathbf{S}$ is a rank $1\leq r\leq K$ matrix, the eigendecomposition of $\mathbf{S}$ gives a set of orthonormal vectors, $\{\mathbf{v}_{i}\}_{i=1}^{r}$, such that for any $\pmb{a} \in \mathbb{R}^{r}$ we have the expansion $\pmb{a}=\sum_{i=1}^{r}\left(\pmb{a}^{\top}\mathbf{v}_{i}\right)\mathbf{v}_{i}$. Notice that the set $\{\mathbf{v}_{i}\}_{i=1}^{r}$ are the right singular vectors of $\mathbfcal{X}$, then it is true that
\begin{equation}
 \mathbfcal{X}\left(\pmb{a}\right)=\sum_{i=1}^{r}\left(\pmb{a}^{\top}\mathbf{v}_{i}\right)\mathbfcal{X}\mathbf{v}_{i}=\sum_{i=1}^{r}\sigma_{i}\left(\pmb{a}^{\top}\mathbf{v}_{i}\right)\boldsymbol{\psi}_{i}.\nonumber 
\end{equation}
\noindent This implies that $\mathbfcal{X}\mathbf{v}_{i}=\sigma_{i}\boldsymbol{\psi}_{i}$ and we have $\boldsymbol{\psi}_{i}=\sigma_{i}^{-1}\mathbfcal{X}\mathbf{v}_{i}$.  Now, suppose that we have some $\mathbf{z}\in \text{sp}\{\boldsymbol{\psi}_{i}\}_{i=1}^{r}$. Then we have the expansion given by $\mathbf{z}=\sum_{i=1}^{r}\innp{\mathbf{z}}{\boldsymbol{\psi}_{i}}_{\mathbb{H}^{L}}\boldsymbol{\psi}_{i}$. By Theorem 7.6 of \cite{weidmann1980}, we have that $\mathbfcal{X}^{*}$ has an SVD with the same eigentriples of $\mathbfcal{X}$ and we obtain the following
\begin{equation}
\mathbfcal{X}^{*}\mathbf{z}=\sum_{i=1}^{r}\innp{\mathbf{z}}{\boldsymbol{\psi}_{i}}_{\mathbb{H}^{L}}\mathbfcal{X}^{*}\boldsymbol{\psi}_{i}=\sum_{i=1}^{r}\sigma_{i}\innp{\mathbf{z}}{\boldsymbol{\psi}_{i}}_{\mathbb{H}^{L}}\mathbf{v}_{i} \nonumber
\end{equation}
\noindent which implies that $\mathbfcal{X}^{*}\boldsymbol{\psi}_{i}=\sigma_{i}\mathbf{v}_{i}$ and we have that $\mathbf{v}_{i}=\sigma_{i}^{-1}\mathbfcal{X}^{*}\boldsymbol{\psi}_{i}$
\end{proof}
\newpage
\begin{proof}[Proof of Lemma 4.1]

\ \\
\begin{itemize}
\item[\textit{i)}] Let $M_{j_{q}}=\sum_{i=0}^{j_{q}}d_{i}$, then we obtain the following elements of $\mathbb{H}_{d}$
\begin{align*}
\vec{\hat{y}}_i^{\left(1\right)}&=\begin{pmatrix}\hat{y}_{i}^{\left(1\right)}& 0 & \cdots & 0 \end{pmatrix}=\sum_{q=1}^{d_{1}}c_{i,\ell_{q}}^{\left(1\right)}\vec{\nu}_{q}\\
\vec{\hat{y}}_i^{\left(2\right)}&=\begin{pmatrix}0 & \hat{y}_{i}^{\left(2\right)} & 0 & \cdots & 0 \end{pmatrix}=\sum_{q=d_1+1}^{d_{1}+d_2}c_{i,\ell_{q}}^{\left(2\right)}\vec{\nu}_{q}\\
&\vdots \\
\vec{\hat{y}}_i^{\left(j_q\right)}&=\begin{pmatrix}0 & \cdots & 0 & \hat{y}_{i}^{\left(j_{q}\right)} & 0 & \cdots & 0 \end{pmatrix}=\sum_{q=M_{j_{q}-1}+1}^{M_{j_{q}}}c_{i,\ell_{q}}^{\left(j_q\right)}\vec{\nu}_{q}\\
&\vdots \\
\vec{\hat{y}}_i^{\left(p\right)}&=\begin{pmatrix}0& \cdots & 0 & \hat{y}_{i}^{\left(p\right)} \end{pmatrix}=\sum_{q=M_{j_{p}-1}+1}^{d}c_{i,\ell_{q}}^{\left(p\right)}\vec{\nu}_{q}.
\end{align*}
\noindent From this, we find that any $\vec{\hat{y}}_{i} \in \mathbb{H}_{d}$ can be expressed as
\begin{align*}
\vec{\hat{y}}_{i}&=\begin{pmatrix}\hat{y}_{i}^{\left(1\right)}& \hat{y}_{i}^{\left(2\right)}& \cdots & \hat{y}_{i}^{\left(j_q\right)} & \cdots & \hat{y}_{i}^{\left(p\right)} \end{pmatrix}\\ &=\vec{\hat{y}}_i^{\left(1\right)}+\vec{\hat{y}}_i^{\left(2\right)}+\cdots+\vec{\hat{y}}_i^{\left(j_{q}\right)}+\cdots+\vec{\hat{y}}_i^{\left(p\right)}=\sum_{q=1}^{d}c_{i,\ell_{q}}^{\left(j_{q}\right)}\vec{\nu}_{q}.
\end{align*}
\item[\textit{ii)}] This part of the proof is a direct consequence of the proof of part \textit{i)}
\end{itemize}
\end{proof}

\begin{proof}[Proof of Lemma 4.2]

The proof of this Lemma is almost identical to the proof of Lemma 4.1 of \cite{haghbin2019} and holds without loss of generality.
\end{proof}

\begin{proof}[Proof of Lemma 4.3]
\ \\
\begin{itemize}
\item[\textit{i)}] Let $M_{j_{q}}=\sum_{i=0}^{j_q} d_{i}$ and denote the $i^{\text{th}}$ element of $\mathbf{b}_{k}$ with $b_{k,i}$, then we obtain the following elements of $\mathbb{H}_{d}^{L}$

\begin{align*}
\pmb{x}_{k}^{\left(1\right)}&=\begin{pmatrix}
\vec{\hat{y}}_{k}^{\left(1\right)}\\
\vec{\hat{y}}_{k+1}^{\left(1\right)}\\
\vdots\\
\vec{\hat{y}}_{k+L-1}^{\left(1\right)}
\end{pmatrix}=\begin{pmatrix}
\vec{\hat{y}}_{k}^{\left(1\right)}\\
0\\
\vdots\\
0
\end{pmatrix}+\begin{pmatrix}
0\\
\vec{\hat{y}}_{k+1}^{\left(1\right)}\\
0\\
\vdots\\
0
\end{pmatrix}+\begin{pmatrix}
0\\
0\\
\vdots\\
0\\
\vec{\hat{y}}_{k+L-1}^{\left(1\right)}
\end{pmatrix}=\sum_{i=1}^{Ld_{1}}b_{k,i}\boldsymbol{\phi}_{i}\\
&=\sum_{q_{i}=1}^{d_{1}}\sum_{r_{i}=1}^{L}c_{k+r_{i}-1,\ell_{q_{i}}}^{\left(1\right)}\boldsymbol{\phi}_{i}\\
\pmb{x}_{k}^{\left(2\right)}&=\begin{pmatrix}
\vec{\hat{y}}_{k}^{\left(2\right)}\\
\vec{\hat{y}}_{k+1}^{\left(2\right)}\\
\vdots\\
\vec{\hat{y}}_{k+L-1}^{\left(2\right)}
\end{pmatrix}=\begin{pmatrix}
\vec{\hat{y}}_{k}^{\left(2\right)}\\
0\\
\vdots\\
0
\end{pmatrix}+\begin{pmatrix}
0\\
\vec{\hat{y}}_{k+1}^{\left(2\right)}\\
0\\
\vdots\\
0
\end{pmatrix}+\begin{pmatrix}
0\\ 0\\ \vdots\\ 0\\
\vec{\hat{y}}_{k+L-1}^{\left(2\right)}
\end{pmatrix}=\sum_{i=Ld_{1}+1}^{L\left(d_{1}+d_{2}\right)}b_{k,i}\boldsymbol{\phi}_{i}\\
&=\sum_{q_{i}=d_{1}+1}^{d_{1}+d_{2}}\sum_{r_{i}=1}^{L}c_{k+r_{i}-1,\ell_{q_{i}}}^{\left(2\right)}\boldsymbol{\phi}_{i}\\
&\vdots
\end{align*}
\begin{align*}
\pmb{x}_{k}^{\left(p\right)}&=\begin{pmatrix}
\vec{\hat{y}}_{k}^{\left(p\right)}\\
\vec{\hat{y}}_{k+1}^{\left(p\right)}\\
\vdots\\
\vec{\hat{y}}_{k+L-1}^{\left(p\right)}
\end{pmatrix}=\begin{pmatrix}
\vec{\hat{y}}_{k}^{\left(p\right)}\\
0\\ \vdots\\ 0
\end{pmatrix}+\begin{pmatrix}
0\\
\vec{\hat{y}}_{k+1}^{\left(p\right)}\\
0\\ \vdots\\ 0
\end{pmatrix}+\begin{pmatrix}
0\\ 0\\ \vdots\\ 0\\
\vec{\hat{y}}_{k+L-1}^{\left(p\right)}
\end{pmatrix}=\sum_{i=LM_{p-1}+1}^{Ld}b_{k,i}\boldsymbol{\phi}_{i}\\
&=\sum_{q_{i}=M_{p-1}+1}^{d}\sum_{r_{i}=1}^{L}c_{k+r_{i}-1,\ell_{q_{i}}}^{\left(p\right)}\boldsymbol{\phi}_{i}.
\end{align*}

As a result, we find that $\pmb{x}_{k}=\pmb{x}_{k}^{\left(1\right)}+\pmb{x}_{k}^{\left(2\right)}+\cdots+\pmb{x}_{k}^{\left(p\right)}=\sum_{i=1}^{Ld}b_{k,i}\boldsymbol{\phi}_{i}$ and the coefficients found in $\mathbf{b}_{k}$ are found in equation (4.3) of the manuscript.
\item[\textit{ii)}]
\begin{equation*}\label{expansion of X}
\mathbfcal{X}(\pmb{a})=\sum_{k=1}^{K}a_{k}\pmb{x}_k
=\sum_{i=1}^{Ld}\left(\sum_{k=1}^{K} b_{k,i}a_{k}\right)\pmb {\phi}_{i}
=\mathbfcal{P}(\mathbf{B}\pmb{a}).
\end{equation*}
\end{itemize}
\end{proof}

\begin{proof}[Proof of Thm. 4.1]

This proof is a direct consequence of Theorem 4.1 of \cite{haghbin2019}
\end{proof}

\begin{proof}[Proof of Thm. 5.1]
\ \\
\begin{itemize}
\item[\textit{i)}] Let $\pmb{a} \in \mathbb{R}^{K}$. Then we have that
\begin{equation}
\mathcal{U}\mathbfcal{X}\left(\pmb{a}\right)=\sum_{k=1}^{K}a_{k}\mathcal{U}\pmb{x}_{k}=\sum_{k=1}^{K}a_{k}\underline{\mathbf{x}}_{k}=\underline{\mathcal{X}}\left(\pmb{a}\right) \nonumber
\end{equation}
\noindent and as such, we have that $\underline{\mathcal{X}}=\mathcal{U}\mathbfcal{X}$.\\
\item[\textit{ii)}] Again, let $\pmb{a} \in \mathbb{R}^{K}$, then we have
\begin{equation}
\mathcal{U}\mathbfcal{X}\left(\pmb{a}\right)=\sum_{i=1}^{r}\sigma_{i}\pmb{a}^{\top}\mathbf{v}_{i}\mathcal{U}\boldsymbol{\psi}_{i}=\sum_{i=1}^{r}\sigma_{i}\pmb{a}^{\top}\mathbf{v}_{i}\underline{\boldsymbol{\psi}}_{i}=\underline{\mathcal{X}}\left(\pmb{a}\right). \nonumber
\end{equation}
\noindent This implies that the $i^{\text{th}}$ eigentriple of $\underline{\mathcal{X}}$ is $\left(\sigma_{i},\mathbf{v}_{i},\underline{\boldsymbol{\psi}}_{i}\right)$ and that $\underline{\boldsymbol{\psi}}_{i}=\mathcal{U}\boldsymbol{\psi}_{i}$.
\end{itemize}
\end{proof}

\end{document}